\documentclass[12pt,fleqn]{article}
\usepackage{amssymb}
\usepackage{graphicx}
\usepackage{bigints}
\usepackage{times}
\usepackage{lineno}
\usepackage{hyperref}
\usepackage{movie15}
\usepackage{media9}
\usepackage[sort&compress,comma]{natbib}
\usepackage{setspace}
\DeclareMathOperator{\bx}{\mathbf{x}}
\DeclareMathOperator{\by}{\mathbf{y}}

\begin{document}
\def\D{\Delta}
\def\d{\delta}
\def\r{\rho}
\def\p{\pi}
\def\a{\alpha}
\def\g{\gamma}
\def\ra{\rightarrow}
\def\s{\sigma}
\def\b{\beta}
\def\e{\epsilon}
\def\G{\Gamma}
\def\om{\omega}
\def\l{\lambda}
\def\f{\phi}
\def\w{\psi}
\def\m{\mu}
\def\t{\tau}
\def\c{\chi}
 \title{}

\title{\sffamily\textbf{Diversity and coevolutionary dynamics in high-dimensional phenotype spaces}}
\vspace{1.5cm}
\author{\vspace*{-0mm} \\
Michael Doebeli$^\ast$ \& Iaroslav Ispolatov$^{\ast\ast}$\\
{\normalsize $^{\ast}$ Departments of Zoology and Mathematics, University
of British Columbia,}
\vspace*{-2mm}\\
{\normalsize 6270 University Boulevard, Vancouver B.C. Canada, V6T 1Z4; doebeli$@$zoology.ubc.ca}\\
\vspace*{1mm}{\normalsize $^{\ast\ast}$ Departamento de Fisica, Universidad de Santiago de Chile}
\vspace*{-2mm}\\
{\normalsize Casilla 302, Correo 2, Santiago, Chile; jaros007$@$gmail.com}
}
\vskip 1cm

\date{\vspace{10mm}\normalsize\today}
\maketitle
\vskip 10mm
\noindent 
\vskip 10 mm
{\bf Supporting Material:} Appendix A
\vskip 10 mm
{\bf RH:} Diversity and coevolutionary dynamics
\noindent 
\vskip 10 mm
{\bf Corresponding author:} Michael Doebeli, Department of Zoology and Department of
  Mathematics,  University of British Columbia, 6270 University Boulevard, Vancouver B.C. Canada, V6T 1Z4, Email: doebeli@zoology.ubc.ca.\\

\newpage
%\linenumbers
\noindent
{\bf \large Abstract}
\vskip 0.3cm
\noindent We study macroevolutionary dynamics by extending microevolutionary competition models to long time scales. It has been shown that for a general class of competition models, gradual evolutionary change in continuous phenotypes (evolutionary dynamics) can be non-stationary and even chaotic when the dimension of the phenotype space in which the evolutionary dynamics unfold is high. It has also been shown that evolutionary diversification can occur along non-equilibrium trajectories in phenotype space. 
We combine these lines of thinking by studying long-term coevolutionary dynamics of emerging lineages in multi-dimensional phenotype spaces. We use a statistical approach to investigate the evolutionary dynamics of many different systems. We find: 1) for a given dimension of phenotype space, the coevolutionary dynamics tends to be fast and non-stationary for an intermediate number of coexisting lineages, but tends to stabilize as the evolving communities reach a saturation level of diversity; and 2) the amount of diversity at the saturation level increases rapidly (exponentially) with the dimension of phenotype space. 
These results have implications for theoretical perspectives on major macroevolutionary patterns such as adaptive radiation, long-term temporal patterns of phenotypic changes, and the evolution of diversity. 

\newpage
\section*{Introduction}

One of the fundamental problems in evolutionary biology is to understand how microevolutionary processes generate macroevolutionary patterns. In particular, the emergence of macroevolutionary changes in the speed of evolution \citep{simpson1944,gould_eldredge1977}, and of macroevolutionary changes in patterns of species diversity \citep{rosenzweig1995, schluter2000} have long been of great interest. For example, \cite{uyeda_etal2011} have recently proposed that over macroevolutionary time scales, relatively short intermittent bursts of high rates of evolutionary change should alternate with long periods of bounded phenotypic fluctuations. 
Also, there is much discussion about whether species diversity saturates over evolutionary time in a given environment \citep{rosenzweig1995, harmon_harrison2015, rabosky_hurlbert2015}. Phylogenetic analysis has been used to shed light on these questions \citep{pennell_harmon2013}, but mechanistic models in which short-term ecological interactions are extrapolated to yield long-term patterns of diversity and evolutionary change have only recently been developed. Most of these models have been used to study the long-term evolution of diversity by analyzing processes of community assembly emerging from short-term ecological dynamics \citep{loeuille_loreau2005,loeuille_loreau2009, ito_dieckmann2007, allhoff_etal2015, rosindell_etal2015, gascuel_etal2015}.
In particular, these papers have mainly focussed on how diversity changes over time, but not on how the nature of the coevolutionary dynamics of a given set of coexisting species changes as the diversity changes. In fact, in all these models, the evolutionary dynamics for a fixed amount of diversity, i.e., for a given set of species, converge to an equilibrium. However, if one wants to understand macroevolutionary changes in the ``tempo and mode'' \citep{simpson1944} of evolution, one not only needs to consider how diversity changes over evolutionary time, but also how such changes in diversity affect the nature of evolutionary dynamics \citep{pennell_harmon2013}. Indeed, there is evidence from evolution experiments with microbes that evolutionary dynamics in more diverse communities are qualitatively different from the evolutionary dynamics in less diverse communities \citep{lawrence_etal2012}. Here we present a theoretical investigation of the questions of how diversity affects the complexity of coevolutionary dynamics.

In general, the number of different phenotypes that affect ecological and evolutionary processes is an important quantity. For example,  determining the dimensionality of niche space in ecological food webs is a classical problem \citep{cohen1977, eklof_etal2013}, and it has recently been shown that including more phenotypic dimensions in models for community assembly has a strong effect on the structure of the emerging food webs \citep{allhoff_etal2015}. Implicitly, the importance of the dimension of phenotype space is also acknowledged in phylogenetic research through the notion of ``adaptive zones'' \citep{simpson1944, uyeda_etal2011}. In particular, it is thought that much of the extant diversity has evolved as a consequence of lineages entering new adaptive zones, which can be interpreted from the phenotypic perspective as an increase in the dimension of phenotype space. In general, given the large number of phenotypic properties that determine an individual's life history and ecology in almost any species, one would expect that ecological interactions are generally determined by many phenotypic properties, and that selection pressures emerging from ecological interactions in turn affect many phenotypes simultaneously. 

For example, comprehensive modelling of the metabolic network in {\it E. coli} cells comprises more than 2000 reactions \citep{yoon_etal2012}. These reactions are in turn controlled by thousands of genes in a complicated interaction network whose exact workings are largely unknown. Nevertheless, many of the genes contributing to this network of metabolic reactions will be under selection in any given environmental setting, and as a consequence, a large number of phenotypic properties have the potential to undergo evolutionary change. It is generally not known how exactly these phenotypic properties impinge on birth and death rates of individual organisms, and hence what exactly the ecological selection pressures are on these properties. Nevertheless, it seems clear that in general, many phenotypes will evolve at the same time, i.e., that evolution generally takes place in high-dimensional phenotype spaces.

We have recently argued that if evolution takes place in high-dimensional phenotype spaces, then the evolutionary dynamics, that is, the phenotypic change over evolutionary time, can be very complicated, i.e., non-stationary and often chaotic \citep{doebeli_ispolatov2014,ispolatov_etal2015}. In low-dimensional phenotype spaces, non-equilibrium evolutionary dynamics are less likely. However, if a species evolving on a simple attractor gives rise to diversification, the effective  dimensionality of the evolving system increases, as the species that emerge from diversification coevolve, driven by both intra- and interspecific ecological interactions. 
Thus the total dimensionality of the resulting dynamical system describing multispecies coevolution is the number of species times the dimensionality of the phenotype space in which each species evolves. Based on our earlier results \citep{doebeli_ispolatov2014,ispolatov_etal2015}, one could then expect that due to the increase in dimensionality, diversification  leads to more complicated evolutionary dynamics in each of the coevolving species. On the other hand, as a multispecies community becomes more diverse and evolves towards saturation, the available niches tend to get filled, and hence evolutionary change has to become highly coordinated between interacting species and thus constrained, potentially leading to simplified evolutionary dynamics. It is thus unclear how the nature of the evolutionary dynamics changes as the pattern of diversity changes during community assembly.

We investigate these issues by applying the framework of adaptive dynamics \citep{geritz_etal1998}  to a general class of competition models. The main question we address is, how does the complexity of long-term coevolutionary dynamics depend on the diversity of the coevolving community? We show that in low-dimensional phenotype spaces, there is a humped-shaped relationship between diversity and the complexity of evolutionary dynamics: in communities with low diversity, coevolutionary dynamics are often simple, i.e., stationary in the long-time limit; for intermediate degrees of diversity, non-stationary (complex) coevolutionary dynamics are common, and each of the species in the community evolves on a complicated trajectory in phenotype space; and for high amounts of diversity, coevolutionary dynamics become simple again, i.e., stationary. In particular, as communities reach diversity saturation, e.g. through adaptive diversification \citep{doebeli2011}, coevolutionary dynamics change from complex to simple. 

Our results are relevant for a number of issues concerning patterns of macroevolution. For example, the results suggest that during processes of adaptive radiation \citep{schluter2000, gavrilets_losos2009, seehausen2015}, evolutionary dynamics are more complicated early in the radiation than late in the radiation, a pattern that corresponds to the ``early-burst'' perspective of macroevolution that has attracted much attention in recent years \citep{gavrilets_losos2009, harmon_etal2010, slater_pennell2013}. Our results also show that the level at which diversity saturates depends on the dimensionality of phenotype space, with higher dimensions allowing for more diversity. This observation is in accordance with data from radiations in fishes \citep{seehausen2015} and points to the possibility of a microevolutionary mechanism for the ``blunderbass theory" of temporal patterns of macroevolutionary changes and diversification \citep{uyeda_etal2011}: if evolution operates on the dimension of phenotype space on a very slow time scale, then on shorter time scales diversity may saturate and thereby generate relatively stationary evolutionary dynamics, whereas on longer time scales the dimension of phenotype space may increase, e.g. due to gene duplications, thus generating a new burst of non-equilibrium (co-)evolutionary dynamics until the diversity reaches a new saturation level. Such patterns of intermittent bursts have recently been found in the phylogenies of birds and echinoids \citep{brusatte_etal2014, hopkins_smith2015}, and the bursts have been attributed to the evolution of flight capabilities and of novel feeding techniques, respectively, both of which can be interpreted as an increase in the dimensionality of the relevant phenotype space. This perspective may also shed light on the question of whether diversity saturates or not \citep{harmon_harrison2015, rabosky_hurlbert2015}: diversity may saturate for a given dimension of phenotype space, but evolutionary innovation in the form of new phenotypic dimensions may intermittently generate room for additional bouts of evolutionary diversification. 

\section*{Methods}
\subsection*{Single-cluster adaptive dynamics}
As in \cite{doebeli_ispolatov2014}, we study a general class of models for frequency-dependent competition in which ecological interactions are determined by $d$-dimensional phenotypes, where $d\geq1$.  For simplicity, we consider homogeneous systems, so no spatial coordinates are included.  The ecological interactions are described by a competition kernel $\alpha(\mathbf{x}, \mathbf{y})$ and by a carrying capacity $K(\mathbf{x})$, where $\mathbf x,\mathbf y\in\mathbb{R}^d$ are the $d$-dimensional continuous phenotypes of competing individuals. The competition kernel $\alpha$  measures the  competitive impact that an individual of phenotype $\mathbf x$ has on an individual of phenotype $\mathbf y$, and we assume that  $\alpha(\mathbf{x}, \mathbf{x})=1$ for all $\mathbf x$. Assuming logistic ecological dynamics, $K(\mathbf{x})$ is then the equilibrium density of a population that is monomorphic for phenotype $\mathbf x$. The adaptive dynamics of the phenotype $\mathbf x$ is a system of differential equations for $d\mathbf x/dt$. To derive the adaptive dynamics, one defines the invasion fitness $f(\mathbf{x}, \mathbf{y})$ as the per capita growth rate of a rare mutant phenotype $\mathbf y$ in the monomorphic resident $\mathbf x$ population that is at its ecological equilibrium $K(\mathbf x)$: 
 
 \begin{align}
 \label{fitness}
f(\mathbf{x}, \mathbf{y}) = 1 - \frac{ \alpha(\mathbf{x}, \mathbf{y}) K(\mathbf{x})}{K(\mathbf{y})}.
\end{align}

\noindent The expression for the invasion fitness reflects the fact that the growth rate of the mutant $\mathbf y$ is negatively affected by the effective density experienced by the mutant, $\alpha(\mathbf{x}, \mathbf{y}) K(\mathbf{x})$, discounted by the carrying capacity $K(\mathbf y)$ of the mutant (see \citep{doebeli2011,doebeli_ispolatov2014} for more details).  Note that $f(\mathbf{x}, \mathbf{x})=0$ for all $\mathbf x$. The invasion fitness $f(\mathbf{x}, \mathbf{y}) $ gives rise to the selection gradients in the $i=1,...,d$ phenotypic components:

\begin{align}
\label{sg}
s_i(\mathbf{x}) \equiv \frac{\partial f(\mathbf{x}, \mathbf{y})}{\partial y_i}{\big |_{\mathbf{y}=\mathbf{x}}} =  - \frac{\partial \alpha(\mathbf{x}, \mathbf{y})}{\partial y_i}{\big |_{\mathbf{y}=\mathbf{x}}} + \frac{\partial K(\mathbf{x})}{\partial x_i}\frac{1}{K(\mathbf{x})},
\end{align}

\noindent The selection gradients in turn define the adaptive dynamics as a system of differential equations on phenotype space $\mathbb{R}^d$, which is given by

 \begin{align}
 \label{AD1}
 \frac{d\mathbf{x}}{dt} = \mathbf M(\mathbf x)\cdot\mathbf{s}(\mathbf{x}).
\end{align} 

\noindent Here $\mathbf{s}(\mathbf{x})$ is the column vector $(s_1(\mathbf x),...,s_d(\mathbf x))$, and $\mathbf M(\mathbf x)$ is the mutational variance-covariance matrix. In this matrix, the diagonal elements contain information about the size and rate of mutations in each of the phenotypic dimensions, whereas the off-diagonal elements contain information about the covariance between mutations in two different phenotypic dimensions. This matrix essentially captures ``evolvability'' of a population and  generally depends on the current resident phenotype $\mathbf x$, and influences the speed and direction of evolution. For simplicity, we assume here that this matrix is the identity matrix. For more details on the derivation of the adaptive dynamics (\ref{AD1}) we refer to a large body of primary literature (e.g. \citep{dieckmann_law1996,geritz_etal1998, diekmann2003, leimar2009, doebeli2011}). We note that the adaptive dynamics (\ref{AD1}) can be derived analytically as a large-population limit  of an underlying stochastic, individual-based model that is again defined based on the competition kernel $\alpha(\mathbf{x}, \mathbf{y})$ and the carrying capacity $K(\mathbf{x})$ \citep{dieckmann_law1996, champagnat_etal2006,champagnat_etal2008}.

Specifically, here we consider a class of systems that are defined by competition kernels of the form 

\begin{align}
\label{comp}
\a(\mathbf{x},\mathbf{y})=\exp\left [\sum_{i,j=1}^d
b_{ij}(x_{i}-y_{i})x_{j}
-\sum_{i=1}^d\frac
{(x_{i}-y_{i})^2}{2\s_i^2}\right].
\end{align}

\noindent Here the coefficients $b_{ij}$ in the first sum on the right hand side are arbitrary and correspond to the simplest form of a generic, non-symmetric competition kernel that can generate non-stationary evolutionary dynamics. It can be interpreted as the lowest-order (non-trivial) term from a Taylor expansion of an unknown non-symmetric competition function. Adaptive dynamics of asymmetric competition has been studied quite extensively (e.g. \cite{law_etal1997,kisdi1999,doebeli2011}), and is necessary to generate single-species non-equilibrium dynamics in high-dimensional phenotype spaces \citep{doebeli_ispolatov2013, doebeli_ispolatov2014}.
The second sum on the right hand side represents ``Gaussian competition'', according to which the competitive impact between individuals increases with phenotypic similarity between the competing individuals. The parameters $\s_i$ measure how fast the effect of competition declines as phenotypic distance in the $i$-component increases. For the carrying capacity we assume  

\begin{align}
\label{K}
K(\mathbf{x})=\exp\left(-\frac{\sum_i^d x_{i}^4}{4}\right).
\end{align}

\noindent This implies that the carrying capacity imposes an element of stabilizing selection for the phenotype $\mathbf x=0$, at which the carrying capacity is maximal. Thus, the frequency-dependent component of selection is generated by the competition kernel, whereas the frequency-independent component of selection is due to the carrying capacity. With these assumptions, the adaptive dynamics (\ref{AD1}) become

\begin{align}
\label{AD2}
\frac{ d x_i}{dt}=\sum_{i=1}^d b_{ij}x_j - x_i^3, \quad i=1,...,d.
\end{align}

\noindent We note that  the terms $-x_i^3$ in (\ref{AD2}) are due to the carrying capacity and serve to contain the trajectories of (\ref{AD2}) in a bounded domain of phenotype space. Also, the Gaussian part of the competition kernel does not affect the adaptive dynamics of monomorphic populations, i.e., the $\s_i$ do not appear in (\ref{AD2}), because the Gaussian part always has a maximum at the current resident, and hence the corresponding first derivative in the selection gradient (\ref{sg}) is 0. 

The system of ODEs  (\ref{AD2}) describes the trajectory of an evolving monomorphic population in phenotype space $\mathbb{R}^d$. In \cite{doebeli_ispolatov2014} we have shown that for general competition kernels $\a$ such trajectories can be very complicated, particularly when the dimension $d$ is large. With complex evolutionary dynamics, trajectories can be quasi-periodic or chaotic, and typically visit many different regions of phenotype space over evolutionary time. When $d$ is low the dynamics tend to be simpler, and often converge to an equilibrium attractor.  We can assess the likelihood of equilibrium dynamics for a given dimension $d$ by choosing the $d^2$ coefficients  $b_{ij}$ in (\ref{AD2}) randomly and independently, e.g. from a normal distribution with mean 0 and variance 1, solving the resulting adaptive dynamics (\ref{AD2}) and checking whether it converges to an equilibrium. If this is done repeatedly, we can approximate the probability of equilibrium dynamics as the fraction of runs that converged to an equilibrium. For $d=1$ the probability of equilibrium dynamics is of course 1, and for $d=2,3,4$, the resulting probabilities of equilibrium dynamics are approximately $85\%$, $81\%$ and $74\%$, respectively. These are the dimensions that we will primarily use in the analysis presented below, but we note that the probability of equilibrium dynamics goes to 0 for large $d$ \citep{edelman1997,doebeli_ispolatov2014}.

\subsection*{Multi-cluster adaptive dynamics}
Here we are interested in the question of how diversification and subsequent coexistence of species (also called phenotypic clusters or simply clusters through the text) affects the evolutionary dynamics. While the Gaussian term in the competition kernel (\ref{comp}) does not affect the adaptive dynamics of single monomorphic populations, this term is crucial for determining whether evolutionary diversification occurs. For one-dimensional phenotype spaces ($d=1$) this is very well known and is encapsulated in the concept of evolutionary branching \citep{metz_etal1992, geritz_etal1998, doebeli2011}. An evolutionary branching point is an equilibrium point of (\ref{AD2}) that is both an attractor for the adaptive dynamics and a fitness minimum. The reason that such points exist in the competition models considered here is precisely that the Gaussian term does not affect the adaptive dynamics, but does affect the curvature of the fitness landscape, i.e., the second derivative of the invasion fitness (\ref{fitness}). In particular, small enough $\s_i$'s in the Gaussian term will make any equilibrium point a fitness minimum, and hence will give rise to evolutionary diversification. Evolutionary branching in scalar traits has been described in a plethora of different models (for an overview we refer to Eva Kisdi's website at the Department of Mathematics and Statistics at the University of Helsinki, http://www.mv.helsinki.fi/home/kisdi/addyn.htm). In high-dimensional phenotype spaces, equilibrium points of (\ref{AD2}) can also be fitness minima along some directions in phenotype space. For this to happen the Hessian matrix of second derivatives of the invasion fitness (\ref{fitness}), evaluated at the equilibrium, must have positive eigenvalues. Indeed, in higher dimensional phenotype spaces the conditions for the existence of positive eigenvalues of this Hessian matrix, and hence for diversification, generally become less stringent \citep{doebeli_ispolatov2010,debarre_etal2014,svardal_etal2014}. 

Importantly, evolutionary diversification can also occur from non-equilibrium adaptive dynamics trajectories \citep{ito_dieckmann2014,ispolatov_etal2016}. If the adaptive dynamics (\ref{AD2}) exhibit non-equilibrium dynamics, the crucial quantity determining whether diversification occurs is again the Hessian matrix of second derivatives of the invasion fitness (\ref{fitness}), but now restricted to the subspace of phenotype space that is orthogonal to the selection gradient \citep{ispolatov_etal2016}. Essentially, diversification can occur in orthogonal directions in which this Hessian has positive curvature, and hence in which the invasion fitness has a minimum. Because the population is still evolving along the selection gradient, elucidating the exact conditions for diversification requires a careful analysis \citep{ito_dieckmann2014}. In the present context, the implication of these results is that, just as with equilibrium adaptive dynamics, diversification can occur along non-equilibrium trajectories of (\ref{AD2}) if the $\s_i$ in the Gaussian term of the competition kernel are small enough, i.e., if the frequency dependence generated by Gaussian competition is strong enough \citep{ispolatov_etal2016}. 

To investigate the process of diversification and the subsequent coevolutionary dynamics, we extend the adaptive dynamics (\ref{AD2}) to several coexisting phenotypic clusters as follows. We assume that an evolving community consists of $m$ monomorphic populations, each given by a phenotype $\mathbf x_r$, $r=1,...,m$, with phenotypic components $x_{ri}$, $i=1,...d$ (where $d$ is the dimension of phenotype space). Let $N_r$ be the population density of cluster $\mathbf x_r$. Then the ecological dynamics of the $m$ clusters are given by the system of logistic differential equations

\begin{align}
\label{logistic_pop}
 \frac{d N_r(t)}{ d t} = N_r( t)\left(\frac{ 1 - \sum_{s=1}^m \a(\mathbf x_s, \mathbf x_r) N_s ( t)}{K(\mathbf x_{r})}\right), \quad r=1,...,m.
\end{align}

\noindent Let $N_r^*$, $r=1,...,m$ denote the equilibrium of system (\ref{logistic_pop}) (more generally, for the purposes of deriving the adaptive dynamics, the quantities $N_r^*$ are suitable time averages of population densities over the ecological attractor of (\ref{logistic_pop}); however, our extensive numerical simulations indicated that (\ref{logistic_pop}) always converges to an equilibrium). Making the traditional adaptive dynamics assumption that ecological dynamics occur on a faster time scale than evolutionary dynamics, we calculate the invasion fitness function in cluster $r$ based on the densities $N_r^*$ of the various clusters: 

\begin{align}
\label{invfit}
 f(\mathbf x_1,...,\mathbf x_m,\mathbf x_r')=1 - \frac{\sum_{s=1}^m \a(\mathbf x_{s},\mathbf x_r')N_{s}^*}{K(\mathbf x_r')}.
\end{align}

\noindent Here $\mathbf x_1,...,\mathbf x_m$ describe the phenotypic state of the resident population, and $\mathbf x_r'$ denotes the mutant trait in cluster $r$, $r=1,...,m$. 

Taking the derivative of (\ref{invfit}) with respect to $\mathbf x_r'$ and
evaluating it at the resident, $\mathbf x_r'=\mathbf x_r$, yields the components of the selection gradient $\mathbf {s}_{r}$ for the cluster $r$ as:

\begin{align}
\label{sgr}
s_{ri}= \sum_{s}N_{s}^*\left(-  \frac{1}{K(\mathbf x_{r})}\left.\frac{\partial \a(\mathbf x_{s},\mathbf x_r')}{\partial
     x_{ri}'}\right|_{\mathbf x_r'=\mathbf x_r} +
 \frac{\a(\mathbf x_s,\mathbf x_{r})}{K^2(\mathbf x_{r})}\frac{\partial
   K(\mathbf x_r)}{\partial x_{ri}}\right), \quad i=1,...,d.
\end{align}

\noindent For coevolutionary adaptive dynamics, one has to take into account that the rate of mutations in each evolving phenotypic cluster is proportional to the current population size of that cluster \citep{dieckmann_law1996}, and hence that the speed of evolution is influenced by the population size. In the single-cluster system such consideration only rescales time without affecting the geometry of the trajectory and thus is usually ignored. However, in the multi-cluster system, instead of assuming that the mutational process is described by the identity matrix as in (\ref{AD1}), we now assume that in each cluster $r$, the mutational variance-covariance matrix $M_r$ is a diagonal matrix with entries $N_r^*$. This generates the following $d\cdot m$ differential equations describing the adaptive dynamics in the coevolving community:

\begin{align}
\label{adode}
 \frac{d x_{ri}}{dt}= N_{r}^* s_{ri}, \; i=1,\ldots,d, \; r=1,\ldots, m.
\end{align}

\noindent 
For the multicluster adaptive dynamics, the equation (\ref{invfit},\ref{sgr},\ref{adode}) replace  their single-cluster analogs (\ref{fitness},\ref{sg},\ref{AD1}).
It is important to note that the Gaussian part of the competition kernel $\a$ not only affects whether diversification occurs, but in contrast to the adaptive dynamics of single monomorphic populations, the Gaussian term will indeed affect the coevolutionary adaptive dynamics (\ref{adode}) of the phenotypic clusters that coexist after diversification has occurred, because it affects both the ecological dynamics (\ref{logistic_pop}) and the selection gradient (\ref{sgr}).

\subsection*{Numerical procedure}
To study diversification and subsequent multi-cluster adaptive dynamics, we
implemented the following iterative numerical scenario:

\vskip 0.5 cm
\noindent {\it Step 1:} Each simulation run is initiated with a randomly generated 
$d \times d$ matrix of the coefficients  $b_{ij}$ for 
the competition kernel (\ref{comp}). The coefficients are drawn from a
Gaussian distribution with zero mean and $d^{-1/2}$ variance. As
explained in \citep{doebeli_ispolatov2014}, this is done to keep the sum of the $d$ terms
$\sum_{j=1}^d b_{ij} x_j$ in (\ref{AD2}) of order  $x_i$, i.e. independent of $d$.
Then a certain number of clusters, given by a parameter $m_0$, each with population size of order $1$, are randomly placed
 near the phenotype $0$, i.e., near the maximum of the carrying capacity.
   
\vskip 0.5 cm
\noindent {\it Step 2:} For a given set of phenotypic clusters, the population dynamics of all clusters is solved using the ecological dynamics (\ref{logistic_pop}). The system of differential equations is integrated using a
  4th-order Runge-Kutta algorithm for $\sim 10^3$ time steps of duration $dt \sim 10^{-2}$ to ensure convergence
  to the equilibrium (or, in case there is no such convergence, to ensure a correct calculation of the time average of the various population
densities). If the population density of a given 
  cluster falls below the threshold $N_{min}\sim10^{-8}$, the
  cluster is eliminated from the system. During the ecological
  dynamics the evolutionary dynamics is frozen and evolutionary
  time does not advance.
  
\vskip 0.5 cm
\noindent {\it Step 3:} After calculating the $N_r^*$, $r=1,...,m$ (where $m$ is the current number of clusters), the adaptive dynamics of the phenotypes of the clusters is advanced via
  (\ref{sgr},\ref{adode}) using a 4th-order Runge-Kutta algorithm with a typical
  time-step $d\t \sim 10^{-2}$, by which the evolutionary time is
  advanced as well. After this evolutionary time step, the ecological dynamics are recalculated, potentially preceded by the following step 4, which is only performed if the corresponding time condition is satisfied. 
  
\vskip 0.5 cm
\noindent {\it Step 4:} The level of diversity, i.e., the number of clusters in the system, is controlled as follows. Each $\t_c$ time units the distances between clusters are assessed.
  If the distance between two or more clusters is below a
  threshold 
$\D x \sim 10^{-3}$, these
  clusters are merged, preserving the total population size of the merged clusters and the position of their centre of mass.  Immediately after this comparison step, the total number of
  clusters is compared to the target number of clusters, which is given by a system parameter $m_{max}$. If the current number of clusters is below $m_{max}$, a new
  cluster is created by randomly picking an existing cluster, 
  splitting it in half and separating the two new clusters in a random direction in phenotype space by
   the distance  of the merging threshold, $\D x$. 
    
\vskip 0.5 cm
\noindent {\it Step 5:} In our simulations, we take measurements at regular time intervals (ranging from $\t_m\sim 1 - 10$ time units). One of the main quantities of interest is the average per capita evolutionary speed $v$ in the evolving community,  which is the average of the norms of the vectors of trait variation (evolution) rates in each cluster, weighted by the cluster
population size,  computed as 
\begin{align}
\label{velocity}
v=\sum_{r=1}^m\frac{N_{r} \sqrt{\sum_{i=1}^d (d x_{ri}/dt)^2}}{\sum_{r=1}^m N_{r}}
\end{align}
\noindent This quantity is a strong indicator of the nature of the evolutionary dynamics of the coevolving system. In particular, our very extensive numerical simulations indicate that when the average speed falls below $10^{-2}$, then the system eventually exhibits equilibrium evolutionary dynamics. In contrast, when the average evolutionary speed remains high, the coevolving system tends to exhibit complicated, non-equilibrium dynamics, with the majority of the clusters exhibiting large fluctuations in phenotype space over evolutionary time. An example of such non-equilibrium coevolution is given in the next section. Other measurements include
  the position and population size of all clusters in the system,
 and the number of ``distinct'' clusters
separated by a ``visible'' distance $\D X=0.1$. These measurements can also be averaged over time. 

\vskip 1cm
For any given simulation run initiated by step 1 above, steps 2-5 were repeated iteratively until a specified final simulation time is reached, or until evolution comes to a halt, which by our definition occurs when the average evolutionary speed falls below a threshold, $v<10^{-4}$. 
Our general approach consisted of simulating many different systems according to the above scheme, and then computing statistical characteristics such as the fraction of runs that result in non-equilibrium dynamics, or the average evolutionary speed as a function of the level of diversity (see Results section). 

One crucial feature of our algorithm is the periodic generation of new clusters in step 4, which mimics diversification events, i.e., evolutionary branching. Diversification is thus modeled by simply adding new phenotypic clusters at certain points in time and close to existing clusters. This mimics the sympatric split of an ancestral lineage. Sympatric diversification is a theoretically robust phenomenon \citep{doebeli2011} and our procedure represents a shortcut for this phenomenon necessitated by computational feasibility. If such splitting is not feasible given the current ecological circumstance, the new cluster will not diverge phenotypically from the ancestor, and hence will be merged again with the ancestor (see below). Alternatively, newly generated clusters may go extinct ecologically. In either case, speciation was not successful. Thus, in our models it is the ecological circumstances that determine whether speciation can occur or not, but the process of speciation itself (i.e., the splitting) is performed in a simplified manner. If speciation is successful and the newly generated clusters diverge and persist ecologically, then diversity has increased (unless other clusters go extinct). We note that by construction, the maximal level of diversity in a given simulation run, i.e., the number of different clusters, cannot exceed the parameter $m_{max}$. Therefore, this parameter allows us to control the level of diversity in a given simulation.

There are in principle other, less artificial ways to model diversification. In particular, stochastic, individual-based based models and partial differential equation models \citep{champagnat_etal2006, champagnat_etal2008} have been used to describe the evolutionary dynamics of phenotype distributions. In such models, diversification is an emergent property that is reflected in the formation of new modes in the evolving phenotype distributions. While these techniques are very useful in general, they are currently not computationally feasible for the statistical approach that we employed here, which requires systematic simulation of many different systems. Also, they would not allow for control of the level of diversity, as the number of phenotypic modes would simply be an emergent property of the evolving system. Nevertheless, we have used these alternative techniques to illustrate the robustness of salient results using particular examples. A more detailed description of these techniques is given in the  Appendix. Another alternative would be to assume that new clusters (species) are assigned phenotypes that are chosen randomly in phenotype space, rather than close to an existing cluster. This could correspond to immigration of new species into an existing community. However, we would not expect this to affect our main results, because with complicated evolutionary dynamics, the initial phenotypic position of a given cluster becomes irrelevant after some time. 

Finally, we note that the merging of clusters (species) is done solely for computational reasons and has no biological meaning (apart from designating organisms that are closely related and phenotypically very close as belonging to the same species). Merging of clusters only occurs shortly after a new cluster is seeded close to an existing one, and only if the new cluster does not diverge from the existing one (i.e., only if the ecological conditions for diversification are not satisfied). If divergence is successful, the clusters will never again get close enough to other clusters to be merged because of the repelling force of frequency-dependent competition. Thus, the only function of merging is to prevent the number of clusters from artificially becoming very large.

\section*{Results}
The parameter that controls the level of diversity in our simulations is $m_{max}$, which is the maximal number of different phenotypic clusters allowed to be present at any point in time in an evolving community (see step 4 in the Methods section). Our first result is obtained by allowing this parameter to be very large, so that we can estimate the number of clusters that eventually coexist by simply running the simulations for a long time and recording the number of clusters at which the diversity equilibrates.  We denote by $M_{\s,d}$ the equilibrium number of clusters for a given phenotypic dimension $d$ and strength of the Gaussian component $\s$ in the competition kernel (\ref{comp}).  We found that such equilibrium level of diversity increases exponentially with  the dimension $d$ of phenotype space, and decreases with the strength $\s$  (Figure 1). 
Here and below we assume for simplicity that the $\s_i$ are the same in all phenotypic directions, $\s_i=\s$ for $i=1,...,d$. In the  Appendix we indicate scaling relationships that hold for $M_{\s,d}$ as functions of the parameters $\s$ and $d$. In general, diversity is only maintained if $\s\lesssim1$, which is roughly the scale of the phenotypic range set by the carrying capacity (\ref{K}). Only if $\s\lesssim1$, the equilibrium level of diversity increases exponentially with increasing dimension of phenotype space, Figure 1. 

\begin{figure*}
\centering
\includegraphics[width=0.5\textwidth]{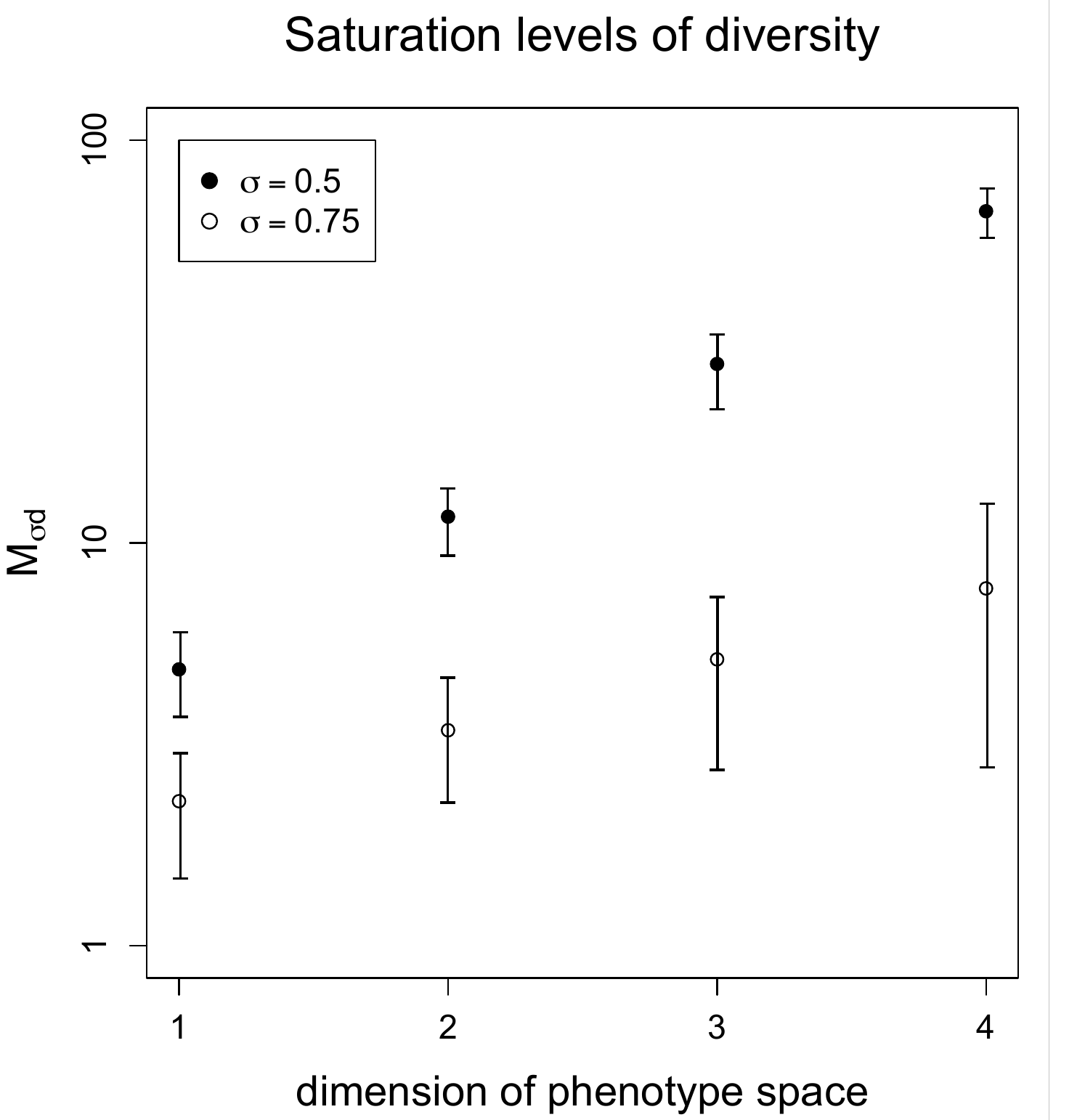}
\caption{Exponential growth of the equilibrium number of clusters $M_{\s,d}$ vs. phenotype dimension $d$ for $\s=0.5$ and $\s=0.75$. For each dimension, $M_{\s,d}$ was determined numerically as an average from several hundred simulation runs in which the parameter $m_{max}$ was set sufficiently high.}
\label{f1}
\end{figure*}

Our main results are now obtained based on the observation that by fixing the parameter $m_{max}$ at a value $\leq M_{\s,d}$ for a given $d$ and $\s$, the community will typically evolve to a diversity level $m_{cluster}$ of approximately $m_{max}$. That is, if the diversity is constrained to be below the maximal level of diversity possible for a given set of parameters, then the diversity will typically evolve to the value set by the constraint. Note that this is an ``average'' statement about many simulations runs, i.e., many different choices of the coefficients $b_{ij}$ and stochastic realizations of cluster splitting. While some simulation runs will result in a diversity that is lower than $m_{max}$ (which may reflect an intrinsic state of the system for the given set of coefficients, or a long-living metastable state which has not yet reached its full diversity), most runs will evolve to the level of diversity that is prescribed by this parameter.

This allows us to then assess, for a given level of diversity, the nature of the coevolutionary dynamics that unfolds in communities with that level of diversity. Two paradigmatic examples are shown in Figure~2. We first set the level of diversity  $m_{max}=12$, which is far below the saturation level $M_{\sigma,d}$ for the given system. Starting from very few clusters the diversity quickly evolves to the level set by $m_{max}$, and the coexisting clusters then exhibit complicated, non-stationary evolutionary dynamics, with all clusters undergoing sustained and irregular fluctuations in phenotype space (Fig.~2a). This type of complicated dynamics is characterized by average evolutionary speeds $v>10^{-2}$. In the same system, but now with a value of $m_{max}$ that lies above the saturation level $M_{\sigma,d}$, the diversity evolves to the saturation level, at which the community consists of ca. 30 coexisting phenotypic clusters (Fig.~2b). In this saturated state, the average evolutionary speed is much lower than $10^{-2}$, and the community exhibits much more stationary coevolutionary dynamics (that would eventually converge to a coevolutionary equilibrium). Moreover, the saturated community exhibits a characteristic pattern of over-dispersion in phenotype space due to competitive repulsion caused by the Gaussian component of the competition kernel (see also Fig. A1 in the Appendix).

To obtain a more systematic characterization of the coevolutionary dynamics as a function of the diversity of the evolving community, we ran, for a given dimension of phenotype space $d$ and strength of competition $\s$, 100 simulations with randomly chosen coefficients $b_{ij}$ for each $m_{max}=1,...,M$, where $M$ is some number that is larger than the saturation level of diversity $M_{\s,d}$. For each run, we recorded the average per capita evolutionary speed $v$ and the number of phenotypic clusters, i.e., the level of diversity, present at the end of $1000$ evolutionary time units (averaged over the last 4 time units). We classified the dynamics into equilibrium dynamics if the average speed $v$ was $<10^{-2}$, and non-equilibrium dynamics otherwise. As mentioned earlier, this was based on individual inspection of many simulation that ran longer than $1000$ time units, which showed that the threshold $10^{-2}$ is a very good indicator of whether the coevolutionary system eventually equilibrates. 

\begin{figure*}
\centering
\includegraphics[width=.9\textwidth]{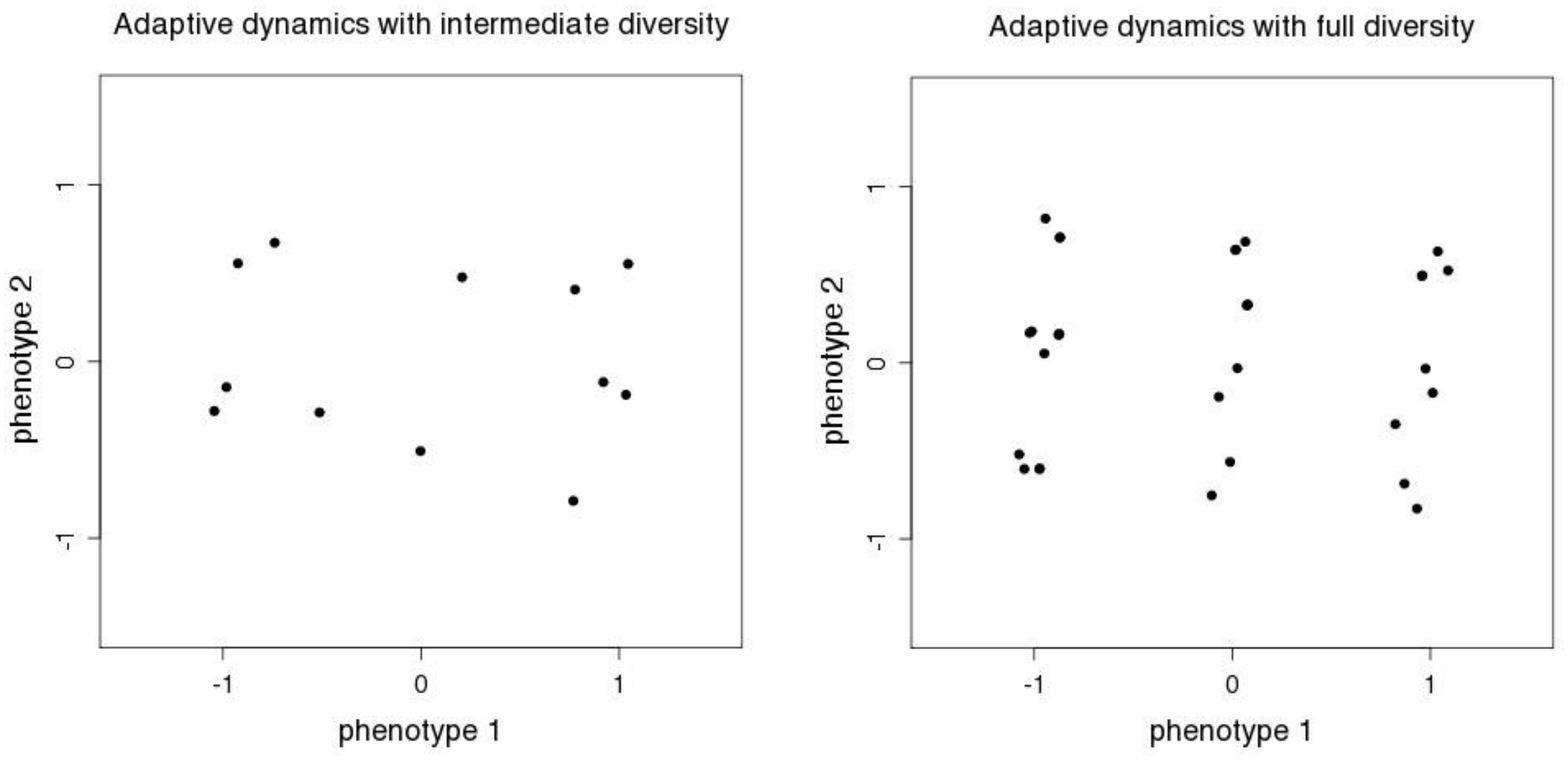}
\includemovie[poster]{7cm}{7cm}{VideoA1.mp4}
\includemovie[poster]{7cm}{7cm}{VideoA2.mp4}
\caption{The panels represent snapshots of the end of the videos and the videos themselves illustrating coevolutionary dynamics with different levels of diversity. In the snapshots and in the videos, phenotypes are projected onto the first two phenotypic dimensions. For these examples, the competition kernel was defined by the coefficients $\{b_{ij}\}$ given in the  Appendix, and $d=3$, $\s=0.5$. In the left panel, $m_{max}=12$, and in the right panel $m_{max}=40$. The saturation level of diversity for this system is $M_{\sigma,d}\sim 30$. With fewer (twelve) clusters, i.e., less diversity (left panel), the coevolutionary dynamics is non-stationary and the clusters undergo sustained and irregular fluctuations in phenotype space, as is apparent in the corresponding video. At the diversity saturation level (right panel), the evolutionary dynamics slows down and becomes almost stationary. The systems were run for 400 time units, and the average evolutionary speed at the end of the simulation runs was $v=3.49\times10^{-2}$ in the first case, and $v=5.54\times10^{-4}$ in the second case.}
\label{m0}
\end{figure*}

Our main results are shown in Figures 3 and 4. The general pattern is that the probability of non-equilibrium dynamics increases as diversity increases from single-cluster communities to communities with a few clusters (Figure 3). 
For intermediate diversity, the fraction of non-equilibrium dynamics remains high. For communities with high diversity, the fraction of non-equilibrium dynamics starts to decrease, and almost almost all communities with a diversity close to the saturation level $M_{\s,d}$  exhibit equilibrium coevolutionary dynamics. 
To illustrates these trends, we  give a more detailed account of the average velocities $v$ defined in (\ref{velocity}) in the coevolving communities (Fig.~4). It shows that there is an exponential decrease in the average speed as the diversity increases, and that there is a substantial fraction of low-diversity communities that exhibit equilibrium dynamics. 
The exact shape of these patterns depends on $d$ and $\s$ (Figures 3 and 4), but whenever diversification is possible, the overall trend is that non-equilibrium dynamics are most likely at intermediate levels of diversity, and that high levels of diversity tend to generate equilibrium coevolutionary dynamics. 

\begin{figure*}
\centering
\includegraphics[width=.9\textwidth]{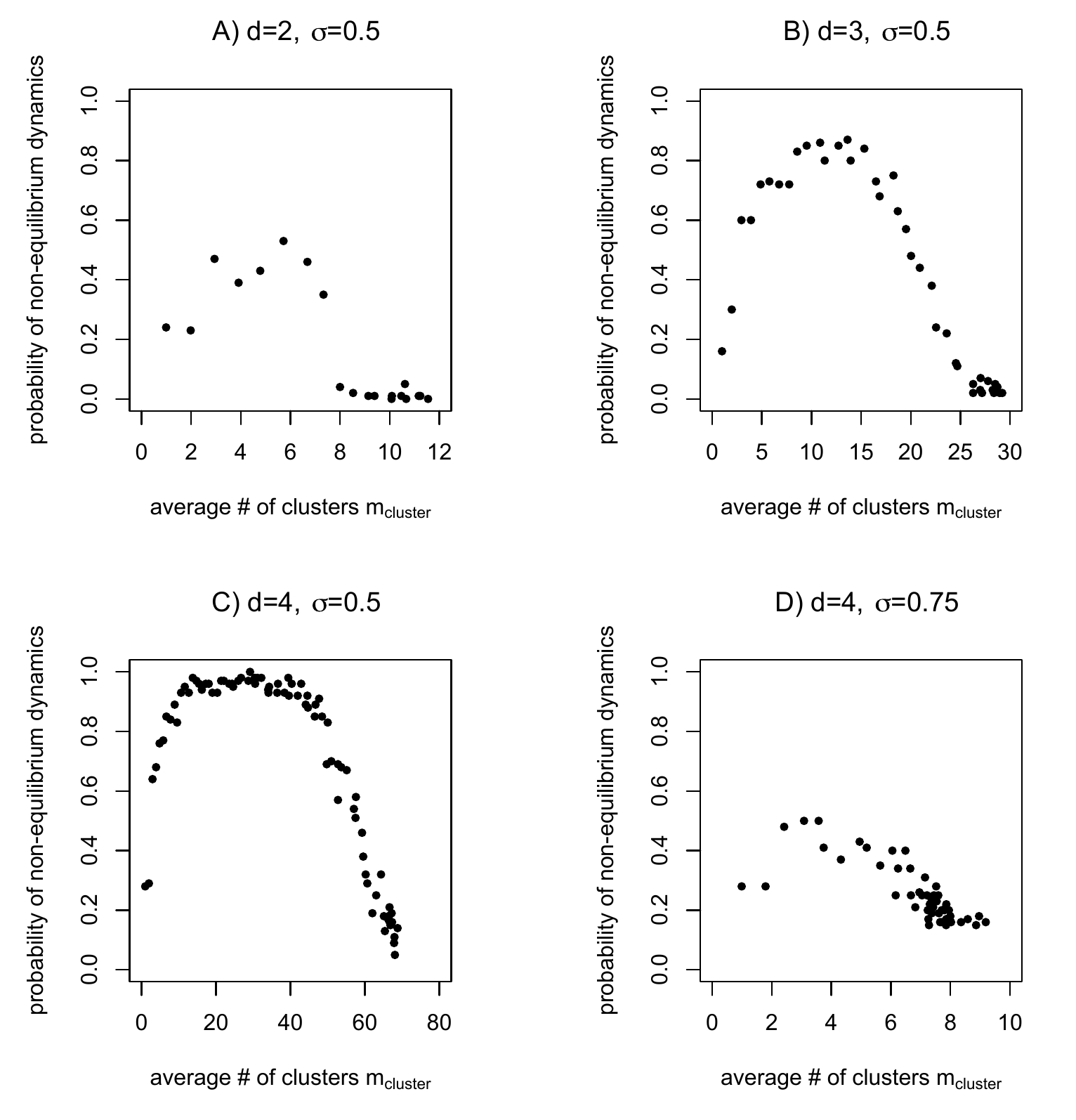}
\caption{The probability of non-equilibrium evolutionary dynamics vs the number of distinct clusters $m_{cluster}$ present in the evolving community for  $d=2, \s=0.5$ (panel A), $d=3, \s=0.5$ (panel B), $d=4, \s=0.5$ (panel C),$d=4, \s=0.75$ (panel D). For each panel, the results were obtained by running, for each integer value of  $m_{max}$ in an interval $[1,M]$ with $M>M_{\sigma,d}$ (the diversity threshold),100 simulations with different random initial conditions and sets of $b_{ij}$ coefficients for 1000 time units. For each of these simulation runs, we recorded the final evolutionary speed and the final number of clusters. We then determined what fraction of the 100 simulation runs for a given $m_{max}$ had an evolutionary speed above the equilibrium dynamics threshold of $10^{-2}$, and then plotted that fraction against the average of the final cluster number calculated from the 100 simulations. }
\label{f2}
\end{figure*}  
\begin{figure*}
\centering
\includegraphics[width=.9\textwidth]{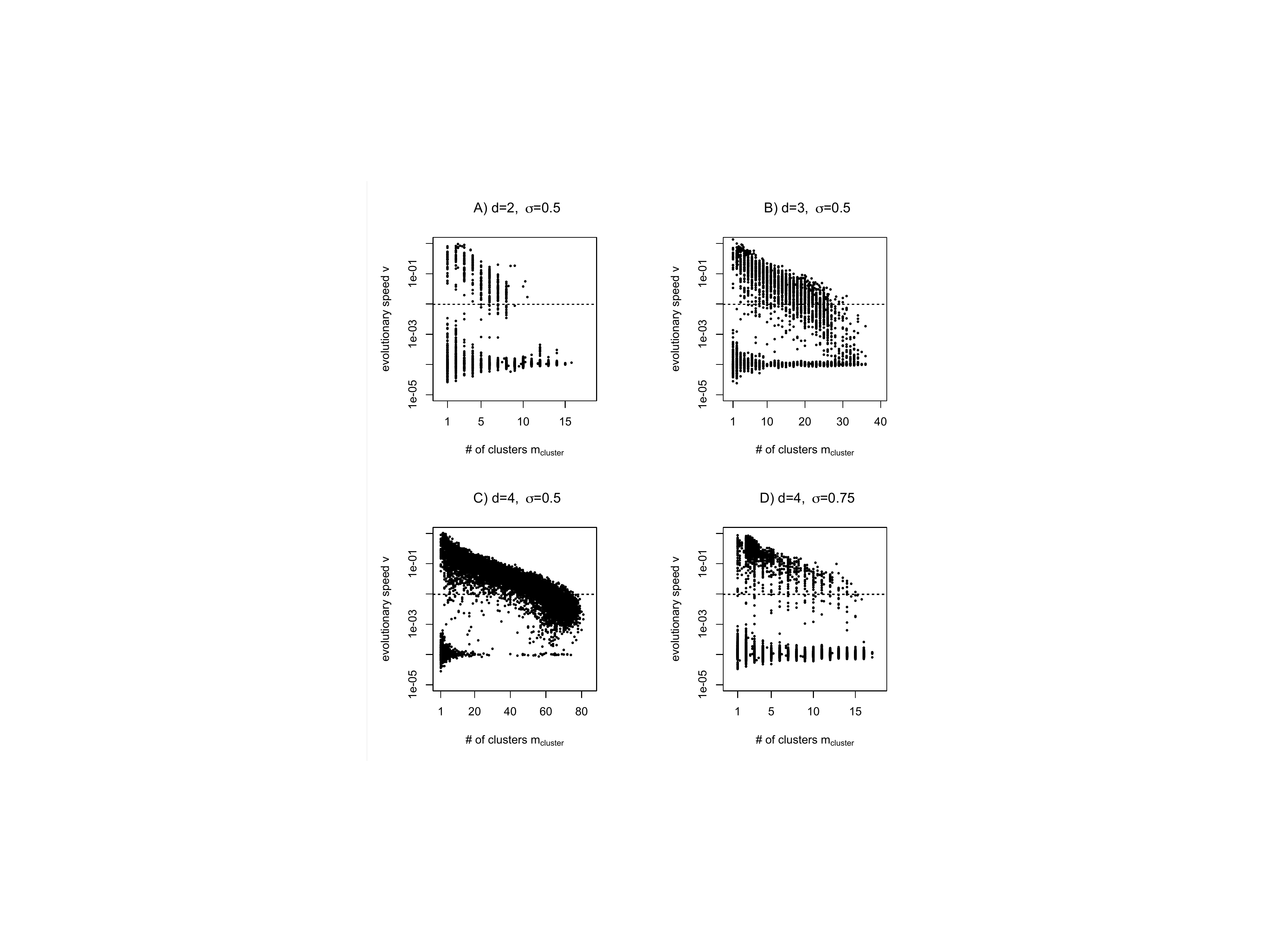}
\caption{The per capita evolutionary speed $v$ vs the number of distinct clusters $m_{cluster}$ for  $d=2, \s=0.5$ (panel A), $d=3, \s=0.5$ (panel B), $d=4, \s=0.5$ (panel C), $d=4, \s=0.75$ (panel D). The points at the bottom of the plots at around $v=10^{-4}$ correspond to systems that have reached the steady state before the final time of the simulation, and the corresponding runs where stopped when the average velocity was below this threshold. For each panel, the results
were obtained from the same simulations runs that were used for Figure 3 by plotting the final
evolutionary speed vs the final cluster number for each simulation run, averaged over a short period at the end of the run (so that the number of clusters is not necessarily an integer). The dashed horizontal lines indicate the threshold for equilibrium evolutionary dynamics (see text).
}
\label{f3}
\end{figure*}

The patterns shown in Figs.~3 and 4 are based on many different simulated communities with different levels of diversity. However, similar patterns can be observed in simulations of single communities as they evolve from low to high diversity, i.e., as they undergo an adaptive radiation. Such a radiation, starting from a single phenotypic cluster, is shown in Fig.~5A. 
Over time the evolving community becomes more diverse due to adaptive diversification, and as a consequence the nature of the coevolutionary dynamics of the community changes. In the example shown in Figure 5A, the coevolutionary dynamics are fast for low to intermediate levels diversity, and then slow down as the community acquires more and more species, until eventually the community reaches a coevolutionary equilibrium at the diversity saturation level. Again, the slowdown of the evolutionary speed during an adaptive radiation appears to occur exponentially with an increase in diversity. This can also be seen by running a given community defined by a given set of coefficients $b_{ij}$ for different values of the parameter $m_{max}$, determining the level of diversity possible in the evolving community. The evolutionary speed exponentially decreases with the diversity given by $m_{max}$ (Fig.~5B). We currently do not have a mechanistic explanation for the exponential decay in evolutionary rates with increasing diversity. It is informative to watch the process of diversification and subsequent evolutionary slowdown unfold dynamically. To verify that the observed dynamical pattern is not an artifact of the adaptive dynamics approximation, we performed the individual-based and partial differential equation simulations of the same system. The movies in Videos in the Appendix, corresponding to the scenario used for Figures~2B and 5A, confirm that all three methods produce qualitatively similar evolutionary pictures. The detailed descriptions of the individual-based and partial differential equation methods are given in the  Appendix. 

Another interesting, although perhaps not so surprising observation for single adaptive radiations concerns the rate of accumulation of new species in the evolving ecosystem. Figure 5C shows the number of species as a function of time during the adaptive radiation scenario used for Figure 5A, illustrating that the rate of diversification is highest at the beginning of the radiation, and then slows down as the community evolves towards the diversity saturation level. The details of these dynamics depend on system parameters, and in particular on the rate at which new species are introduced into the system, but the qualitative behaviour of diversification rates, which are initially high and then slow down, is common to all adaptive radiations generated by our models.

\begin{figure*}
\centering
\includegraphics[width=0.8\textwidth]{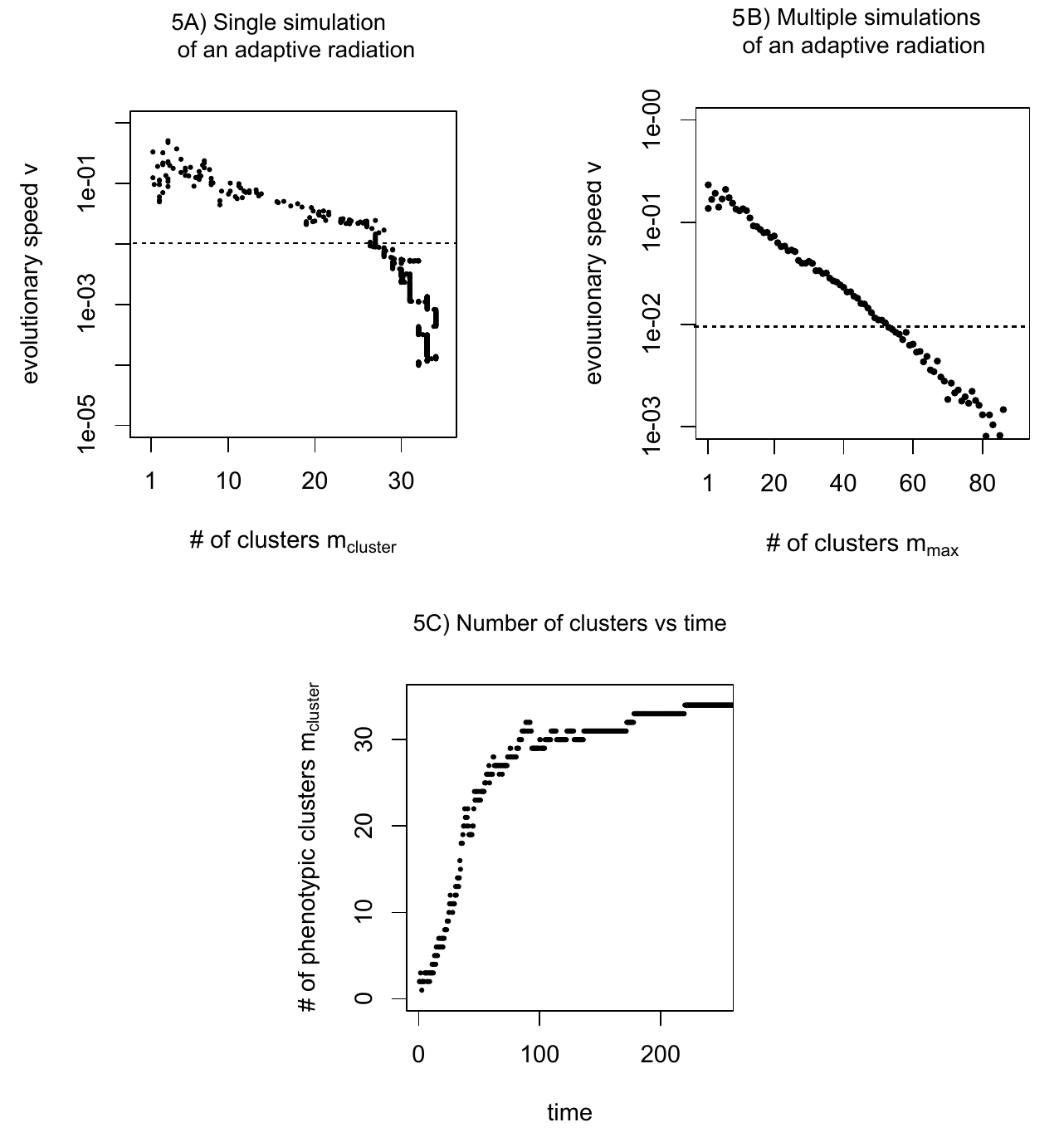}
\caption{The per capita evolutionary speed $v$ vs the number of distinct clusters $m_{cluster}$ present in the evolving community for a single simulation run, corresponding to a single adaptive radiation. This is the same scenario as that shown in Figure 2B: the competition kernel was defined by the coefficients $\{b_{ij}\}$ given in the Appendix, and $d=3$, $\s=0.5$, and $m_{max}=40$. The system exhibits equilibrium single-cluster dynamics. Note the initial increase of the speed $v$ as the system evolves from a single-cluster to a few-cluster regime, followed by a steady exponential decay of $v$ until the evolving community is diverse enough to exhibit equilibrium dynamics. 5B: The per capita evolutionary speed $v$ as a function of the maximum number of clusters $m_{max}$ for a choice of coefficients $\{b_i\}$ for $d=4$ and $\s=0.5$. The coefficients $b_{ij}$ are given in the  Appendix. 
The system exhibits non-equilibrium single-cluster dynamics. For each $m_{max}$ the system was run for 1000 time units and the per capita velocity $v$ was averaged over the last 200 time steps. Note that while the case shown in 5B corresponds to a single choice of the coefficient matrix $b_{ij}$, separate simulations were run for different values of $m_{max}$. This is in contrast to 5A, which shows data from a single simulation run (with a large $m_{max}$). The dashed horizontal lines indicate the threshold for equilibrium evolutionary dynamics (see text). 5C: Number of distinct phenotypic clusters (species) as a function of time for the scenario shown in 5A. The rate of diversification is high at the beginning of the radiation and slows down as the evolving community reaches saturation diversity.}
\label{f4}
\end{figure*}

\section*{Discussion}

We investigated the expected long-term evolutionary dynamics resulting from competition for resources in models for gradual evolution in high-dimensional phenotype spaces. In reality, most organisms have many different phenotypic properties that impinge on their ecological interactions in generally complicated ways, and here we assumed that multi-dimensional phenotypes determine logistic ecological dynamics through the competition kernel and the carrying capacity. We then used a coevolutionary adaptive dynamics algorithm to extend the ecological dynamics to macroevolutionary time scales, and we used a statistical approach to capture general properties of the ensuing evolutionary dynamics. 
 
If the negative frequency-dependence generated by the competition kernel is strong enough, competition results in repeated adaptive diversification, and hence in communities of coevolving phenotypic species. By randomly choosing many different competition kernels, we showed that the complexity of the coevolutionary dynamics in such communities is expected to be highest for intermediate levels of phenotypic diversity. In particular, as the evolving communities increase in diversity towards the saturation level, i.e., the maximal number of different species that can coexist, the evolutionary dynamics becomes simpler, and communities at the saturation level are expected to exhibit a coevolutionary equilibrium. We also showed that the diversity saturation level increases exponentially with the dimension of phenotype space.

We have used a statistical approach to determine the expected long-term evolutionary dynamics resulting from competition for resources. We have assumed that multi-dimensional phenotypes determine logistic ecological dynamics through the competition kernel and the carrying capacity, and we then used a coevolutionary adaptive dynamics algorithm to extend the ecological dynamics to macroevolutionary time scales. 
If the negative frequency-dependence generated by the competition kernel is strong enough, competition results in repeated adaptive diversification, and hence in communities of coevolving phenotypic species. By randomly choosing many different competition kernels, we showed that the complexity of the coevolutionary dynamics in such communities is expected to be highest for intermediate levels of phenotypic diversity. In particular, as the evolving communities increase in diversity towards the saturation level, i.e., the maximal number of different species that can coexist, the evolutionary dynamics becomes simpler, and communities at the saturation level are expected to exhibit a coevolutionary equilibrium. We also showed that the diversity saturation level increases exponentially with the dimension of phenotype space. 

Our interpretation of these findings is that in low-dimensional phenotype spaces such as the ones considered here, evolutionary dynamics of single species are expected to converge to an equilibrium \citep{doebeli_ispolatov2014}. However, as diversity increases, the different phenotypic clusters will ``push'' each other around evolutionarily due to frequency-dependent competition. This occurs mostly due to the repulsive nature of pairwise interaction induced by the Gaussian term in the competition kernel (\ref{comp}): clusters that move further apart decrease competition felt from each other. For example,  a splitting of a cluster stuck in an attractive fixed point of the adaptive dynamics creates two offspring which may become moving again if the repulsion between clusters is stronger than the attraction of the fixed point.  As long as diversity is not very high, i.e., as long as there is enough available niche or unoccupied phenotype space, this typically results in non-equilibrium coevolutionary dynamics, thus leading to an increase in evolutionary complexity with phenotypic diversity. As the diversity keeps increasing towards saturation levels, which for each phenotypic dimension is  determined roughly by the ratio of the widths of the carrying capacity and the competition kernel (see Video 2), the available carrying capacity niche gets filled, so that the evolving clusters ``have nowhere to go'' evolutionarily. An analogy with gas-liquid-solid phase transitions may illustrate this in the following way: As in the dynamics of molecules, the adaptive dynamics of phenotypic clusters contains a pairwise-repulsive term, which originates from the Gaussian term in the competition kernel. A few-cluster regime qualitatively corresponds to the gas phase, when the range of the repulsive interaction is significantly less than the typical distance between clusters. As the number and thus density of clusters increases, the repulsive interaction becomes more relevant, constraining the individual motion of clusters and resulting in a liquid-like behaviour, where clusters are predominantly localized and occasionally hop to a new location. Finally, the maximum cluster density creates a crystal-like structure, albeit not necessarily entirely symmetric due to the randomly generated $b_{ij}$ terms in the adaptive dynamics.  The motion of individual clusters is heavily constrained by its neighbours via mutual repulsion, while the collective motion of an ensemble of clusters is limited by the carrying capacity function.  
Thus, phenotypic saturation leads to a state in which the coevolving clusters are strongly constrained evolutionarily by the other clusters in the community, and hence to coevolutionary equilibrium dynamics. 

Some empirical support for an initial increase in the complexity of evolutionary dynamics with the number of species in an ecosystem comes from the laboratory evolution experiments of \cite{lawrence_etal2012}, who showed that the speed of adaptation to novel environments is higher in bacterial species that are part of microbial communities with a small number of competitors than when evolving in monoculture. However, our results are seemingly in contrast to previous theoretical results about the effect of diversity on evolutionary dynamics \citep{johansson2008, demazancourt_etal2008}. These authors essentially argued that while a single species is free to evolve in response to changes in the environment, evolution of the same species is more constrained in a community of competitors, in which other species are more likely to evolutionarily occupy new niches. Hence diversity is expected to slow down evolution.  However, these models only describe evolution in 1-dimensional phenotypes, and may thus miss the complexity arising in higher-dimensional spaces. Moreover, even in higher-dimensional spaces, the arguments for evolutionary slowdown presented in \citep{johansson2008, demazancourt_etal2008} essentially correspond to our observation of a slow-down when diversity reaches saturation, at which point evolutionary change in each species is indeed constrained due to competing species occupying all available niches. Our approach also needs to be distinguished from approaches based primarily on ecological dynamics, as in \cite{shtilerman_etal2015}. In these approaches, emerging ecological communities are also modelled by periodically adding new species, but there is no underlying phenotype space that would determine competitive interactions. Instead, every time a new species added, its interaction coefficients with the already existing species are chosen according to a specific, randomized procedure. This leads to interesting results, such as saturating levels of diversity after initially fast and fluctuating increases from low levels of diversity. However, since there is no underlying phenotype space, this approach does not reveal the evolutionary dynamics of continuous phenotypes, and in particular, it does not yield any information about the effects of the dimension of phenotype space on the evolutionary dynamics or on the amount of diversity at saturation.

There has been much interest in recent years in determining the effects of phylogenetic relationships on the functioning of ecosystems (e.g. \cite{ives_godfray2006,cadotte_etal2013,nuismer_harmon2015}). The intuitive notion is that phylogenetic information has predictive power for ecological interactions if recently diverged species are more likely to interact than those that diverged long ago. More specifically, \cite{nuismer_harmon2015} have argued that phylogenetic information is most likely to be relevant for ecosystem dynamics if ecological interactions are based on phenotypic matching, so that species with more similar trait values are more likely to interact strongly. Our models have a component of phenotypic matching due to the Gaussian part of the competition kernel, but they also have a strong component of different types of interactions due to the ``random'' part of the competition kernel given by the coefficients $b_{ij}$. As we have shown, it is this non-Gaussian part of the competition kernel that causes the complicated coevolutionary dynamics, and it is this complexity in turn that makes phylogenetic signal largely irrelevant in our models. 

A full phylogenetic analysis of the macroevolutionary dynamics generated by our models is beyond the scope of this work, but we can provide some basic insights based on the complicated evolutionary dynamics in phenotype space that the different phenotypic clusters (species) perform when there is an intermediate number of clusters in the coevolving community. An example of this is shown in the movie in Figure A1A. Here, after an initial phase of diversification, the community contains 12 coevolving clusters. These clusters move on a complicated evolutionary trajectory, with each cluster undergoing large evolutionary changes without further diversification. No matter what the phylogenetic relationship between these clusters (as given by their emergence from the single initial cluster), it is clear that because of the large evolutionary fluctuations in phenotype space of each cluster (species), there will be no consistent correlation between phylogenetic relationship and phenotypic distance. Even if there were such a correlation (positive or negative) at a particular point in time, it would change over time due to the large evolutionary fluctuations of each cluster over time. This is illustrated in Figure A1B, which shows  that no persistent correlation pattern between phylogenetic and phenotypic distance should be expected in communities with an intermediate amount of diversity. In particular, recently diverged species are not more likely to interact than those diverged less recently, because the evolving community has a short ``phenotypic memory'' due to complicated evolutionary dynamics. 

However, when further diversification is allowed, so that the system reaches its saturation level of diversity, the coevolving community not only becomes more diverse, but the evolutionary dynamics slows down, leading to ever smaller phenotypic fluctuations. In particular, new clusters emerging towards the end of the assembly of the evolutionarily stable community will stay phenotypically closer to their phylogenetically most closely related clusters, i.e., to their parent or sister species. Therefore, in the last phase of community assembly a positive correlation between phylogenetic and phenotypic distance can be expected to build up at least to some extent. This is illustrated in Figure A1B. Thus, weak phylogenetic signals are expected to develop towards the end of community assembly.

Regarding adaptive radiations, two observations emerge from our models. The first concerns the classical notion that rates of diversification should decline over the course of a radiation \citep{schluter2000, gavrilets_losos2009}, a pattern that seems to have good empirical support \citep{schluter2000, mcpeek2008,rabosky_lovette2008, gavrilets_losos2009}. Our models confirm this pattern of declining rates of diversification (Figure 5).  The second observation is that rates of evolution should generally slow down with an increase in diversity. This should not only be true when different ecosystems are compared (Figures 3,4), but also during an adaptive radiation in a single evolving community (Figure 5). Thus, we would expect the evolutionary dynamics to be faster and more complicated early in an adaptive radiation, and to slow down and eventually equilibrate late in the radiation. This corresponds to the so-called ``early-burst'' model of macroevolution \citep{gavrilets_losos2009,harmon_etal2010} in the context of adaptive radiations. This model predicts that when lineages enter novel ``adaptive zones''  \citep{simpson1944}, such as novel ecological niches, evolutionary rates in the lineage should be fast initially and then slow down as the adaptive zone gets filled with diverse phenotypes. \cite{harmon_etal2010} found little evidence for the early-burst model when analyzing a large set of data from many different clades. Nevertheless, these authors noted that younger clades have higher rates of evolution than older clades, which points to the fact that evolutionary rates may slow down with clade age. Moreover, few clades in their data set correspond to the type of very fast adaptive radiation envisaged and observed in our models, and they did not consider high-dimensional phenotypes. Finally, \cite{harmon_etal2010} note that groups with a larger proportion of sympatric species early in their history would be more likely to exhibit an early-burst pattern. In our models, adaptive radiations occur in complete sympatry and indeed produce the early burst pattern. 

According to \cite{slater_pennell2013}, the jury on early-burst models is still out, and in fact substantial evidence for this model has accumulated in recent years. For example, \cite{cooper_purvis2010} reported an early burst in body size evolution in mammals, \cite{weir_mursleen2013} observed an early-burst pattern in the evolution of bill shape during adaptive radiation in seabirds,  \cite{gonzalezvoyer_kolm2011} and \cite{arbour_lopezfernandez2013} reported early-burst patterns in morphological and functional evolution in cichlids, and \cite{benson_etal2014} described patterns of early bursts in the evolution of dinosaur morphology. 

\cite{uyeda_etal2011} have incorporated the early-burst concept into a macroevolutionary perspective in which over very long evolutionary time scales, rare but substantial phenotypic bursts alternate with more stationary periods of bounded phenotypic fluctuations, somewhat  reminiscent of the concept of punctuated equilibrium \citep{gould_eldredge1977} when applied to rates of phenotypic evolution \citep{pennell_etal2014}. We think that the models presented here could provide a microevolutionary basis for such a perspective if they are extended by considering evolutionary change in the dimension of the phenotype space that determines ecological interactions. Such an extended theory would have three time scales: a short, ecological time scale, an intermediate time scale at which co-evolution and single diversifications take place in a given phenotype space, and a long time scale at which the number of phenotypic components increases (or decreases). Our hypothesis would then be that in such systems, periods of bounded evolutionary fluctuations near diversity saturation levels for a given dimension of phenotype space would alternate with bursts of rapid evolutionary change, brought about by an evolutionary increase in phenotypic dimensions and the subsequent increase in diversity and acceleration in evolutionary rates until a new saturation level is reached. The resulting long-term evolutionary dynamics would thus show periods of relative phenotypic stasis alternating with periods of fast evolution. This picture would fit very well with the ``blunderbass'' pattern envisaged in \cite{uyeda_etal2011}. These authors proposed that the intermittent bursts in evolutionary rates are caused by lineages encountering novel ``adaptive zones'' \citep{simpson1944}. Novel adaptive zones would correspond to the opening up of new habitats or new resources, which would in turn correspond to new phenotypes that determine use of the novel adaptive zone. Alternatively, novel adaptive zones could also be generated by the emergence of novel sets of regulatory mechanisms allowing novel uses of already existing habitats and resources (as e.g. when a trade-off constraint is overcome through gene duplication). In either case, novel adaptive zones would correspond to an increase in the dimensionality of ecologically important phenotypes. 

It is interesting to note that such intermittent burst patterns have in fact been observed in phylogenetic data, and that they seem to be connected to novel, ecologically important phenotypes. \cite{hopkins_smith2015} have shown that evolutionary rates in echinoids reveal at least two instances of rapidly accelerating and subsequently declining evolutionary rates, i.e., two intermittent bursts. Moreover, these bursts appear to be associated with the evolution of novel feeding strategies \citep{hopkins_smith2015,slater2015}. Also, \cite{brusatte_etal2014} have shown that an evolutionary burst occurs in the dinosaur-bird transition, and it is tempting to conjecture that this burst was caused by the increase in phenotype dimensionality due to the proliferation of flight capabilities.

There is also good empirical support for our finding that the level at which diversity saturates increases with the dimension of phenotype space. \cite{seehausen2015} has argued that essentially, the high number of  different ecologically relevant traits is the basis for the spectacular radiations of cichlids in African lakes. In conjunction with ecological opportunity, genetic and phenotypic flexibility, which appears to be at least in part due to gene duplications, has allowed this group of fish to reach a much higher diversity than other groups, such as cichlids in rivers or whitefish in arctic lakes, in which fewer phenotypes appear to be ecologically relevant  \citep{hudson_etal2011,seehausen2015}. In this context, we note that incorporating the evolution of the dimension of phenotype space may also shed light on the ongoing debate about whether diversity saturates over evolutionary time or not \citep{harmon_harrison2015, rabosky_hurlbert2015}. It seems that the answer could be ``yes and no'': diversity saturates in the intermediate term for a given dimension of phenotype space, but does not saturate in the long term if the dimension of phenotype space increases over long evolutionary time scales, thus generating recurrent increases in saturation levels.

Our study has a number of limitations that should be addressed in future research. It is currently impractical to perform the statistical analysis presented here for phenotype spaces with dimensions higher than $4$ due to computational limitations. Our results indicate that the diversity saturation level, i.e., the maximal number of coexisting phenotypic clusters, increases rapidly with the dimension $d$ of phenotype space, which makes simulations of communities at saturation levels unfeasible. Nevertheless, we expect the salient result that coevolutionary dynamics slow down as communities reach the saturation level to be true in any dimension as long as the Gaussian component of competition in (\ref{comp}) affects all phenotypic directions. Also, in our approach we have assumed that the phenotypes determining competitive interactions are the same for intra- and inter-specific competition. This may be a fair assumption for closely related species, such as those coevolving in an adaptive radiation. However, for competition in more general ecosystems it may also be relevant to assume that from a total set of $d$ phenotypes, different subsets determine competition within a species and competition with various other species. In addition, to describe general ecosystems and food webs, it will be important to include not just competitive interactions, but also predator-prey and mutualistic interactions, each again determined by potentially high-dimensional phenotypes. Also, throughout we have assumed a simple unimodal form of the carrying capacity to represent the external environment. More complicated forms of the carrying capacity, and hence of the external fitness landscape will likely generate even richer patterns of coevolutionary dynamics and diversification. Finally, we have assumed throughout that evolving populations are well-mixed, and it will be interesting so see how the results generalize to spatially structured ecosystems. All these extensions remain to be developed. 

We are of course aware of the fact that we did not include genetic mixing due to sexual reproduction in our models, and our method of describing diversification by simply adding new phenotypic clusters, although fairly standard, does not take into account the actual process of speciation. In sexual populations, adaptive diversification due to disruptive selection, as envisioned here, requires assortative mating, and the conditions for the evolution of various types of assortative mating, as well as for the likelihood of speciation once assortment is present, have been studied extensively (e.g. \citep{doebeli2011}). A general, if crude conclusion from this work is that when there is enough disruptive selection for diversification to occur in asexual models, then it is likely that adaptive speciation also occurs in the corresponding sexual models, although factors such as the strength of assortment, population size and linkage disequilibrium may become important. It would in principle be possible to incorporate sexual reproduction into the models presented here, e.g. along the lines of \cite{gascuel_etal2015}. Our previous results \citep{doebeli_ispolatov2010, ispolatov_etal2016} indicate that adaptive diversification is generally more likely in high-dimensional phenotype spaces, and we think that the present models serve well as a first approximation to study adaptive diversification and coevolutionary dynamics in evolving communities.

Ultimately, the applicability and relevance of our models for understanding macroevolutionary patterns in nature depends in part on being able to determine evolutionary rates of high-dimensional phenotypes from phylogenetic data, which appears to be a difficult problem \citep{denton_adams2015,adams2013, adams2014}. Nevertheless, overall we think that our approach of incorporating microevolutionary processes based on ecological interactions in high-dimensional phenotype spaces into statistical models for macroevolutionary dynamics has the potential to shed new light on a number of fundamental conceptual questions in evolutionary biology.

\section*{Acknowledgments}
M. D. was supported by NSERC (Canada). I. I. was supported by FONDECYT grant 1151524 (Chile). Both authors contributed equally to this work. 

\newpage
\section*{Correlation between phylogenetic and phenotypic distance}
For each pair of clusters (species) in an evolving community we define the phylogenic distance between them, $Pg$, as the number of links in the path between them on the phylogenic tree. To measure this distance, we add the following scheme to our evolutionary algorithm:
 \begin{itemize}
 \item The system is initialized with a single cluster.
 \item Each cluster splitting event produces two offspring separated by the distance 2. The distance between an offspring and all its existing neighbours is incremented by one.
 \item When two recently split cluster that failed to diverge are merged, the distance between the newly produced common cluster and each of its neighbours is calculated as the minimum of the distances of the two merged clusters minus one. This reflects the observation that merging events only happen with newly split clusters. 
 \end{itemize}
 As a result, at any given time we know phylogenic distances between all pairs of clusters currently present in the system. To quantify the relation between the phenotypic and phylogenic similarity, we compute the correlation $C$ between phylogenetic and phenotypic distance as follows:
\begin{align}
\label{ppcorr}
C=\frac{\langle [Pg - \langle Pg \rangle ][X - \langle X \rangle  ] \rangle}{\sigma_{Pg} \sigma_X},
\end{align}

where $Ph$ and $X$ are phylogenic and phenotypic distances between clusters, $\langle \ldots \rangle$ define the average over all pairs of clusters present in the system and $\sigma_{Pg}$ and $\sigma_X$ are the standard deviations of distances.

The above scheme allows us to track the correlation between phylogenetic and phenotypic distance over time, as illustrated in Figure A1. Fig. A1A shows the time dependence of $C$ for the simulation shown in Video 1, and in Fig. A1B shows the time dependence of $C$ for the simulation shown in Video 2. During the early phase of community assembly the correlation $C$ rapidly decays due to complicated coevolutionary dynamics of the emerging clusters. When the diversity of the coevolving community is kept intermediate (by setting the parameter $m_C$ to intermediate values, as in Video 1), the correlation between phylogenetic and phenotypic distance itself undergoes fluctuations around 0 (Fig. A1A). This is because the clusters in the community with intermediate diversity undergo large phenotypic fluctuations while their phylogenetic relationship is constant, because no further diversification (or extinction) occurs. However, when the diversity is allowed to reach saturation levels (by setting $m_C$ to a large value, as in Video 2), a positive correlation between phylogenetic and phenotypic distance develops in the final stages of community assembly, i.e., as the coevolving community reaches the saturation diversity and hence undergoes much smaller phenotypic fluctuations (Fig. A1B). Note that the correlation is still close to 0 during the early stages of community assembly, but some correlation remains at the end due to clusters emerging in the last phase of community assembly, which tend to stay phenotypically closer to their sister species because evolutionary dynamics become slow and stable.

\setcounter{figure}{0}  

\begin{figure*}
\centering
\includegraphics[width=0.95\textwidth]{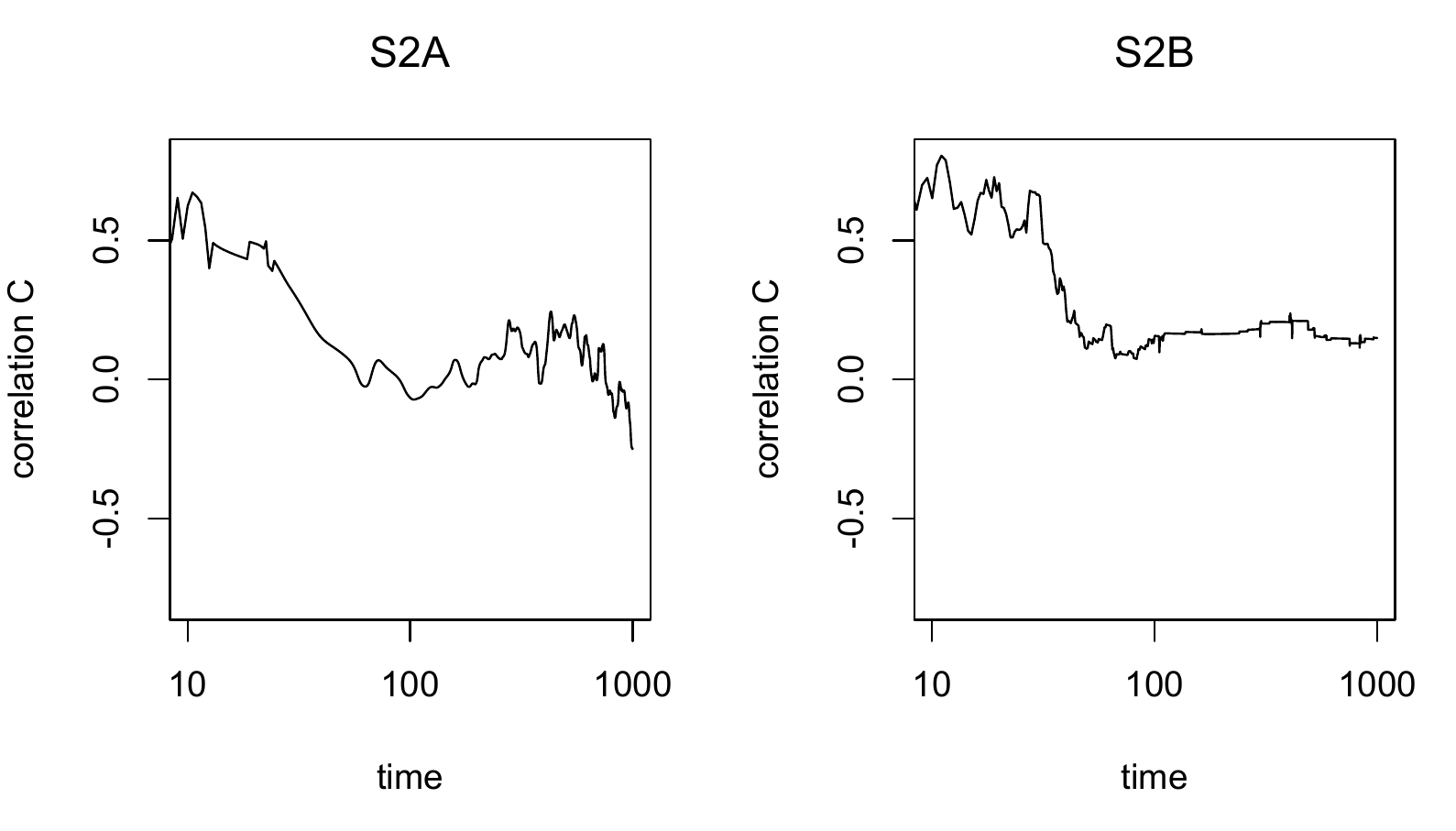}
\renewcommand{\figurename}{Figure A}
\caption{Correlation between phylogenetic and phenotypic distance in an evolving community as a function of time. A1A) corresponds to the simulation run shown in Video A1, and  A1B) corresponds to the simulation run shown in Video A2. In A1A), the correlation decays and then fluctuates around 0, reflecting the fact that diversity is kept intermediate, leading to complicated evolutionary dynamics, during which individual clusters (species) undergo large phenotypic fluctuations, while the phylogenetic relationships are constant once the maximal allowed level of diversity is reached. In A1B), the community is allowed to reach saturation level of diversity. During the later stages of community assembly, the evolutionary dynamics slow down and become more stable, allowing for a positive correlation between phylogenetic and phenotypic distance in the youngest species of the community. Note the the log scale on the time axis.}
\label{f1A}
\end{figure*}

\section* {Individual-based simulations}
Individual-based realizations of the model
were based on the Gillespie algorithm \citep{gillespie1976}
and consisted of the following steps:
\begin{enumerate}
\item  The system is initialized by creating a set of $K_0 \sim 10^3 - 10^4$ individuals with
  phenotypes  $\bx_{k}\in\mathbf{R}^d$ localized around the initial position $\bx_0$
  with a small random spread $|\bx_{k} - \bx_0|\sim10^{-3}$.
\item  Each individual $k$ has a 
constant reproduction rate $\r_{k}=1$ and a death rate 

\noindent $\d_{k}=\sum_{ l \neq k}
A(\bx_{l},\bx{_k})/[K_0K(\bx_{k})]$, as defined by the logistic ecological dynamics.
\item  The total update rate is
given by the sum of all individual rates, $U=\sum_{k}
(\r_{k}+\d_{k})$. 
\item The running time $t$ is incremented by a random number
$\D t$ drawn from the exponential distribution $P(\D t)= U \exp (-\D t  U)$.
\item A particular birth or death event is randomly chosen  with
probability equal to the rate of this event divided by the total update
rate $U$. If  a reproduction event is chosen, the phenotype of an
offspring is offset from the parental phenotype by a 
small mutation randomly drawn from a uniform distribution with
amplitude $\m = 10^{-3} - 10^{-2}$.
\item The individual death rates $\d_{k}$ and the total update rate
$U$ are updated  to take into account the addition or removal of an
individual.
\item  Steps 4-6 are repeated until $t$ reaches a specified end time. 
\end {enumerate}

\noindent The movie in Video A2 shows the dynamics of the individual-based model corresponding to the adaptive dynamics simulation shown in Video A1, which is the same as the scenario used for Video 2 in the main text  (note that the movie in Video A1 runs for $t=1200$ time units, whereas the movie in Video 2 runs for $t=400$ time units).
\newpage
\setcounter{figure}{0}  
\begin{figure*}[h!]
\centering
\includemovie[poster]{7cm}{7cm}{VideoA3.mp4}
\renewcommand{\figurename}{Video A}
\caption{Video of an adaptive radiation. This is the scenario from which data for Figure 5A was extracted. The corresponding coefficients $b_{in}$ for the competition kernel are given in Section 5. Phenotypes are projected onto the first two phenotypic dimensions and shown for three different simulation methods that give qualitatively similar results. This video shows the same multicluster adaptive dynamics simulation as in Video 2, but for a longer time span, $t=0,...,1200$.}
\label{m1}
\end{figure*}
\vskip 1cm
\begin{figure*}[h!]
\centering
\includemovie[poster]{7cm}{7cm}{VideoA4.mp4}
\renewcommand{\figurename}{Video A}
\caption{Video of an adaptive radiation. Phenotypes are projected onto the first two phenotypic dimensions and shown for three different simulation methods that give qualitatively similar results. The video shows an individual-based simulation corresponding to the system shown in Video A1. Details of the individual-based simulations are described in Section 3 of the  Appendix). }
\label{m1i}
\end{figure*}

\vskip 1 cm
\begin{figure*}[h!]
\centering
\includemovie[poster]{7cm}{7cm}{VideoA5.mp4}
\renewcommand{\figurename}{Video A}
\caption{Video of an adaptive radiation. The video shows a numerical simulation of the partial differential equation model explain ins Section 4 of the Appendix. The scenario shown corresponds to the ones shown in Videos A1 and A2. Shown are the projections of the positions of the maxima of the phenotype distribution in the evolving community. Note that the completely deterministic partial differential equation model is symmetric in that it is invariant under the coordinate inversion $\bx\rightarrow - \bx$. Therefore, the solution illustrated in the video is symmetric as well.}
\label{m1pde}
\end{figure*}

\newpage
\section*{Partial differential equation models}
A deterministic large-population limit of the individual-based model is obtained as the partial differential equation (PDE)

\begin{align}
\label{logistic}
 \frac{\partial N(\bx, t)}{\partial t} = N(\bx, t)\left(\frac{ 1 - \int \alpha(\by, \bx) N (\by, t) dy}{K(\bx)}\right)+D\sum_{i=1}^d \frac{\partial^2 N(\bx, t)}{\partial x_i^2},
\end{align}

\noindent where  $N(\bx, t)$ is the population distribution at time $t$ \citep{champagnat_etal2006}. The second term of the right hand side is a diffusion term that describes mutations, 
with the diffusion coefficient typically set to $D\sim 10^{-4} - 10^{-3}$. Local maxima of the solution $N(x,t)$ can be interpreted as positions of the centers of the phenotypic clusters. Their dynamics are shown in Video A3.  
For any given scenario, the corresponding adaptive dynamics solution can be used to determine the single- or few-cluster trajectory, and hence to approximately determine the region occupied by the system in phenotype space over time. Note that the deterministic PDE model is invariant with regard to the coordinate change $\bx\rightarrow -\bx$, and hence its solutions must be symmetric with regard to simultaneous reflection on all coordinate axes. To numerically solve the PDE model (\ref{logistic})
 we  chose a lattice noticeably  
larger than the corresponding adaptive dynamics attractor. The number of bins $B$
in each dimension of this lattice is strongly constrained by memory limitations: An
efficient implementation requires computing and storing an array of
$B^{d} \times B^d$ values of the competition kernel $\alpha(\by_i, \bx_j)$ for the pairwise interactions between
all pairs of sites $i$ and $j$.  With $B=25 -30$ to achieve a reasonable
spatial resolution, the memory constraint makes the
PDE implementation feasible only for $d=2,3$. 

The movie in Video A3 shows the dynamics of the partial differential equation model corresponding to the scenarios shown in Videos A1 and A2.

\section*{Scaling relationship for the diversity at saturation}
The number of clusters at the diversity saturation level, $M_{\s,d}$, can be estimated to be proportional to the volume of the available phenotype space with the linear dimension $L$, divided by the volume occupied by each cluster, which has a typical linear size $\s$:

\begin{align}
\label{volume}
M_{\s_a,d}\approx C_{\s} \frac{L^d} {\s^d}.
\end{align}

\noindent Hence, the following scaling  relationships hold:

\begin{align}
\label{saturation}
M_{\s_a,d}=M_{\s_b,d}\left(\frac{\s_b}{\s_a}\right)^d \quad \text{and }\quad  M_{\s,d_1}=M_{\s,d_2}^{d_1/d_2},
\end{align}
\noindent where $\s_a$ and $\s_b$ denote different strengths of competition, and $C_{\s}$ is a constant of order 1 that takes into account the ``imperfect packing'' occurring when $\s$ and $L$ have similar magnitude. 
Based on this, the equilibrium level of diversity is expected to increase exponentially with increasing dimension of phenotype space (as illustrated Figure 1), and with increasing frequency-dependence (i.e., decreasing $\s$). In general, diversity is only maintained if $\s\lesssim1$, which is roughly the scale of the phenotypic range set by the carrying capacity given by eq. (5) in the main text. 

\section{Specific sets of coefficients used}

The following set of coefficients $b_{ij}$ determining the competition kernel were used for Figures~5A in the main text and for the movies.

\begin{align}
 \label{b_coeff}
\begin{pmatrix} 
   0.407 &  0.498 &  0.287\\
  -0.199 & -1.102 & -0.305\\
   1.387  &-0.896  & 0.341
\end{pmatrix}
\end{align}

The following set of coefficients $b_{ij}$ determining the competition kernel were used for Figure~5B in the main text:

\begin{align}
\label{b_coeff4d}
\begin{pmatrix} 
    -1.289  &   0.682   &  0.217  &  -0.093\\
    -0.223  &  -0.035   &  0.697  &  -0.117\\
    -0.563  &   0.434   & -0.953  &  -0.198\\
     0.119  &   0.398   &  0.183  &   0.530
\end{pmatrix}
\end{align}
%\newpage
\vskip 2cm

\newpage
\bibliography{EvolutionofDiversity}

\begin{thebibliography}{63}
\providecommand{\natexlab}[1]{#1}
\providecommand{\url}[1]{\texttt{#1}}
\providecommand{\urlprefix}{URL }

\bibitem[{Adams(2013)}]{adams2013}
Adams, D.~C., 2013.
\newblock Quantifying and comparing phylogenetic evolutionary rates for shape
  and other high-dimensional phenotypic data.
\newblock Systematic Biology 63:166--177.

\bibitem[{Adams(2014)}]{adams2014}
---{}---{}---, 2014.
\newblock A method for assessing phylogenetic least squares models for shape
  and other high-dimensional multivariate data.
\newblock Evolution 68:2675--2688.

\bibitem[{Allhoff et~al.(2015)Allhoff, Ritterskamp, Rall, Drossel, and
  Guill}]{allhoff_etal2015}
Allhoff, K.~T., D.~Ritterskamp, B.~C. Rall, B.~Drossel, and C.~Guill, 2015.
\newblock Evolutionary food web model based on body masses gives realistic
  networks with permanent species turnover.
\newblock Scientific Reports 5:10955.
\newblock 10.1038/srep10955.

\bibitem[{Arbour and L\'opez-Fern\'andez(2013)}]{arbour_lopezfernandez2013}
Arbour, J.~H. and H.~L\'opez-Fern\'andez, 2013.
\newblock Ecological variation in south american geophagine cichlids arose
  during an early burst of adaptive morphological and functional evolution.
\newblock Proceedings of the Royal Society, London B 280:20130849.
\newblock 10.1098/rspb.2013.0849.

\bibitem[{Benson et~al.(2014)Benson, Campione, Carrano, Mannion, Sullivan,
  Upchurch, and Evans}]{benson_etal2014}
Benson, R. B.~J., N.~E. Campione, M.~T. Carrano, P.~D. Mannion, C.~Sullivan,
  P.~Upchurch, and D.~C. Evans, 2014.
\newblock Rates of dinosaur body mass evolution indicate 170 million years of
  sustained ecological innovation on the avian stem lineage.
\newblock PLoS Biology e1001853.
\newblock 10.1371/journal.pbio.1001853.

\bibitem[{Brusatte et~al.(2014)Brusatte, Lloyd, Wang, and
  Norell}]{brusatte_etal2014}
Brusatte, S.~L., G.~T. Lloyd, S.~C. Wang, and M.~A. Norell, 2014.
\newblock Gradual assembly of avian body plan culminated in rapid rates of
  evolution across the dinosaur-bird transition.
\newblock Current Biology 24:2386--2392.

\bibitem[{Cadotte et~al.(2013)Cadotte, Albert, and Walker}]{cadotte_etal2013}
Cadotte, M., C.~Albert, and S.~Walker, 2013.
\newblock The ecology of differences: assessing community assembly with trait
  and evolutionary distances.
\newblock Ecology Letters 16:1234--1244.

\bibitem[{Champagnat et~al.(2006)Champagnat, Ferri\`ere, and
  Meleard}]{champagnat_etal2006}
Champagnat, N., R.~Ferri\`ere, and S.~Meleard, 2006.
\newblock Unifying evolutionary dynamics: From individual stochastic processes
  to macroscopic models.
\newblock Theoretical Population Biology 69:297--321.

\bibitem[{Champagnat et~al.(2008)Champagnat, Ferri{\`{e}}re, and
  M{\'{e}}l{\'{e}}ard}]{champagnat_etal2008}
Champagnat, N., R.~Ferri{\`{e}}re, and S.~M{\'{e}}l{\'{e}}ard, 2008.
\newblock {From Individual Stochastic Processes to Macroscopic Models in
  Adaptive Evolution}.
\newblock Stochastic Models 24:2--44.
\newblock \urlprefix\url{http://dx.doi.org/10.1080/15326340802437710}.

\bibitem[{Cohen(1977)}]{cohen1977}
Cohen, J.~E., 1977.
\newblock Food webs and the dimensionality of trophic niche space.
\newblock Proceedings of the National Academy of Sciences, USA 74:4533--4536.

\bibitem[{Cooper and Purvis(2010)}]{cooper_purvis2010}
Cooper, N. and A.~Purvis, 2010.
\newblock Density-dependent diversification in north american wood warblers.
\newblock The American Naturalist 175:727--738.

\bibitem[{D{\'{e}}barre et~al.(2014)D{\'{e}}barre, Nuismer, and
  Doebeli}]{debarre_etal2014}
D{\'{e}}barre, F., S.~L. Nuismer, and M.~Doebeli, 2014.
\newblock {Multidimensional (Co)Evolutionary Stability}.
\newblock The American Naturalist 184:158--171.
\newblock \urlprefix\url{http://dx.doi.org/10.1086/677137}.

\bibitem[{Denton and Adams(2015)}]{denton_adams2015}
Denton, J. S.~S. and D.~C. Adams, 2015.
\newblock A new phylogenetic test for comparing multiple high-dimensional
  evolutionary rates suggests interplay of evolutionary rates and modularity in
  lanternfishes (myctophiformes; myctophidae).
\newblock Evolution 69:2425--2440.

\bibitem[{Dieckmann and Law(1996)}]{dieckmann_law1996}
Dieckmann, U. and R.~Law, 1996.
\newblock The dynamical theory of coevolution: A derivation from stochastic
  ecological processes.
\newblock Journal of Mathematical Biology 34:579--612.

\bibitem[{Diekmann(2004)}]{diekmann2003}
Diekmann, O., 2004.
\newblock A beginner's guide to adaptive dynamics.
\newblock Banach Center Publ. 63:47--86.

\bibitem[{Doebeli(2011)}]{doebeli2011}
Doebeli, M., 2011.
\newblock Adaptive diversification.
\newblock Princeton University Press, Princeton.

\bibitem[{Doebeli and Ispolatov(2013)}]{doebeli_ispolatov2013}
Doebeli, M. and I.~Ispolatov, 2013.
\newblock Symmetric competition as a general model for single-species adaptive
  dynamics.
\newblock Journal of Mathematical Biology Pp. 169--184.

\bibitem[{Doebeli and Ispolatov(2014)}]{doebeli_ispolatov2014}
---{}---{}---, 2014.
\newblock {Chaos and unpredictability in evolution}.
\newblock Evolution 68:1365--1373.
\newblock \urlprefix\url{http://dx.doi.org/10.1111/evo.12354}.

\bibitem[{Doebeli and Ispolatov(2010)}]{doebeli_ispolatov2010}
Doebeli, M. and Y.~Ispolatov, 2010.
\newblock Complexity and diversity.
\newblock Science 328:493--497.

\bibitem[{Edelman(1997)}]{edelman1997}
Edelman, A., 1997.
\newblock The probability that a random real gaussian matrix has $k$ real
  eigenvalues, related distributions, and the circular law.
\newblock Journal of Multivariate Analyis 60:203--232.

\bibitem[{Ekl\"of et~al.(2013)Ekl\"of, Jacob, Kopp, Bosch, Castro-Urgal,
  Chacoff, Dalsgaard, de~Sassi, Galetti, Guimar{\~{a}}es, Lom{\'{a}}scolo,
  Gonz{\'{a}}lez, Pizo, Rader, Rodrigo, Tylianakis, V{\'{a}}zquez, and
  Allesina}]{eklof_etal2013}
Ekl\"of, A., U.~Jacob, J.~Kopp, J.~Bosch, R.~Castro-Urgal, N.~P. Chacoff,
  B.~Dalsgaard, C.~de~Sassi, M.~Galetti, P.~R. Guimar{\~{a}}es, S.~B.
  Lom{\'{a}}scolo, A.~M.~M. Gonz{\'{a}}lez, M.~A. Pizo, R.~Rader, A.~Rodrigo,
  J.~M. Tylianakis, D.~P. V{\'{a}}zquez, and S.~Allesina, 2013.
\newblock The dimensionality of ecological networks.
\newblock Ecology Letters 16:577--583.
\newblock \urlprefix\url{http://dx.doi.org/10.1111/ele.12081}.

\bibitem[{Gascuel et~al.(2015)Gascuel, Ferri\`ere, Aguil\'ee, and
  Lambert}]{gascuel_etal2015}
Gascuel, F., R.~Ferri\`ere, R.~Aguil\'ee, and A.~Lambert, 2015.
\newblock {How Ecology and Landscape Dynamics Shape Phylogenetic Trees}.
\newblock Systematic Biology 64:509--607.
\newblock \urlprefix\url{http://dx.doi.org/10.1093/sysbio/syv014}.

\bibitem[{Gavrilets and Losos(2009)}]{gavrilets_losos2009}
Gavrilets, S. and J.~B. Losos, 2009.
\newblock Adaptive radiation: Contrasting theory with data.
\newblock Science Pp. 732--737.

\bibitem[{Geritz et~al.(1998)Geritz, Kisdi, Mesz\'ena, and
  Metz}]{geritz_etal1998}
Geritz, S. A.~H., E.~Kisdi, G.~Mesz\'ena, and J.~A.~J. Metz, 1998.
\newblock Evolutionarily singular strategies and the adaptive growth and
  branching of the evolutionary tree.
\newblock Evolutionary Ecology 12:35--57.

\bibitem[{Gillespie(1976)}]{gillespie1976}
Gillespie, D.~T., 1976.
\newblock A general method for numerically simulating the stochastic time
  evolution of coupled chemical reactions.
\newblock Journal of Computational Physics 42:403--434.

\bibitem[{Gonzalez-Voyer and Kolm(2011)}]{gonzalezvoyer_kolm2011}
Gonzalez-Voyer, A. and N.~Kolm, 2011.
\newblock Rates of phenotypic evolution of ecological characters and sexual
  traits during the tanganyikan cichlid adaptive radiation.
\newblock Journal of Evolutionary Biology 24:2378--2388.

\bibitem[{Gould and Eldredge(1977)}]{gould_eldredge1977}
Gould, S.~J. and N.~Eldredge, 1977.
\newblock Punctuated equilibria: the tempo and mode of evolution reconsidered.
\newblock Paleobiology 3:115--151.

\bibitem[{Harmon and Harrison(2015)}]{harmon_harrison2015}
Harmon, L.~J. and S.~Harrison, 2015.
\newblock Species diversity is dynamic and unbounded at local and continental
  scales.
\newblock The American Naturalist 185:584--593.
\newblock \urlprefix\url{http://dx.doi.org/10.1086/680859}.

\bibitem[{Harmon et~al.(2010)Harmon, Losos, Davies, Gillespie, Gittleman,
  Jennings, Kozak, McPeek, Moreno-Roark, Near, Purvis, Ricklefs, Schluter,
  Schulte, Seehausen, Sidlauskas, Torres-Carvajal, Weir, and
  Mooers}]{harmon_etal2010}
Harmon, L.~J., J.~B. Losos, T.~J. Davies, R.~G. Gillespie, J.~L. Gittleman,
  W.~B. Jennings, K.~H. Kozak, M.~A. McPeek, F.~Moreno-Roark, T.~J. Near,
  A.~Purvis, R.~E. Ricklefs, D.~Schluter, J.~A. Schulte, O.~Seehausen, B.~L.
  Sidlauskas, O.~Torres-Carvajal, J.~T. Weir, and A.~O. Mooers, 2010.
\newblock Early bursts of body size and shape evolution are rare in comparative
  data.
\newblock Evolution 64:2385--2396.

\bibitem[{Hopkins and Smith(2015)}]{hopkins_smith2015}
Hopkins, M.~J. and A.~B. Smith, 2015.
\newblock Dynamic evolutionary change in post-paleozoic echinoids and the
  importance of scale when interpreting changes in rates of evolution.
\newblock Proceedings of the National Academy of Sciences, USA 112:3758--3763.

\bibitem[{Hudson et~al.(2011)Hudson, Vonlanthen, and
  Seehausen}]{hudson_etal2011}
Hudson, A.~G., P.~Vonlanthen, and O.~Seehausen, 2011.
\newblock Rapid parallel adaptive radiations from a single hybridogenic
  ancestral population.
\newblock Proceedings of the Royal Society, London B 278:58--66.

\bibitem[{Ispolatov et~al.(2016)Ispolatov, Madhok, and
  Doebeli}]{ispolatov_etal2016}
Ispolatov, I., V.~Madhok, and M.~Doebeli, 2016.
\newblock Individual-based models for adaptive diversification in
  high-dimensional phenotype spaces.
\newblock Journal of Theoretical Biology 390:97--105.

\bibitem[{Ispolatov et~al.(2015)Ispolatov, Madhok, Allende, and
  Doebeli}]{ispolatov_etal2015}
Ispolatov, J., V.~Madhok, S.~Allende, and M.~Doebeli, 2015.
\newblock Chaos in high-dimensional dissipative dynamical systems.
\newblock Scientific Reports 5:12506.
\newblock 10.1038/srep12506.

\bibitem[{Ito and Dieckmann(2007)}]{ito_dieckmann2007}
Ito, H.~C. and U.~Dieckmann, 2007.
\newblock A new mechanism for recurrent adaptive radiations.
\newblock American Naturalist 170:E96--E111.

\bibitem[{Ito and Dieckmann(2014)}]{ito_dieckmann2014}
---{}---{}---, 2014.
\newblock Evolutionary branching under slow directional evolution.
\newblock Journal of Theoretical Biology 360:290--314.
\newblock \urlprefix\url{http://dx.doi.org/10.1016/j.jtbi.2013.08.028}.

\bibitem[{Ives and Godfray(2006)}]{ives_godfray2006}
Ives, A.~R. and H.~C.~J. Godfray, 2006.
\newblock Phylogenetic analysis of trophic associations.
\newblock The American Naturalist 186:E1--E14.

\bibitem[{Johansson(2008)}]{johansson2008}
Johansson, J., 2008.
\newblock Evolutionary responses to environmental changes: how does competition
  affect adaptation?
\newblock Evolution 62:421--435.

\bibitem[{Kisdi(1999)}]{kisdi1999}
Kisdi, E., 1999.
\newblock Evolutionary branching under asymmetric competition.
\newblock Journal of Theoretical Biology 197:149--162.

\bibitem[{Law et~al.(1997)Law, Marrow, and Dieckmann}]{law_etal1997}
Law, R., P.~Marrow, and U.~Dieckmann, 1997.
\newblock On evolution under asymmetric competition.
\newblock Evolutionary Ecology 11:485--501.

\bibitem[{Lawrence et~al.(2012)Lawrence, Fiegna, Behrends, Bundy, Phillimore,
  Bell, and Barraclough}]{lawrence_etal2012}
Lawrence, D., F.~Fiegna, V.~Behrends, J.~Bundy, A.~Phillimore, T.~Bell, and
  T.~Barraclough, 2012.
\newblock Species interactions alter evolutionary responses to a novel
  environment.
\newblock PLOS BIOLOGY 10.
\newblock \urlprefix\url{http://dx.doi.org/10.1371/journal.pbio.1001330}.

\bibitem[{Leimar(2009)}]{leimar2009}
Leimar, O., 2009.
\newblock Multidimensional convergence stability.
\newblock Evolutionary Ecology Research 11:191--208.

\bibitem[{Loeuille and Loreau(2005)}]{loeuille_loreau2005}
Loeuille, N. and M.~Loreau, 2005.
\newblock Evolutionary emergence of size-structured food webs.
\newblock Proceedings of the National Academy of Sciences, USA 102:5761--5766.

\bibitem[{Loeuille and Loreau(2009)}]{loeuille_loreau2009}
---{}---{}---, 2009.
\newblock Emergence of complex food web structure in community evolution
  models.
\newblock Pp. 163--178, \emph{in} Community Ecology. Oxford University Press,
  Oxford.
\newblock
  \urlprefix\url{http://dx.doi.org/10.1093/acprof:oso/9780199228973.003.00013}.

\bibitem[{de~Mazancourt et~al.(2008)de~Mazancourt, Johnson, and
  Barraclough}]{demazancourt_etal2008}
de~Mazancourt, C., E.~Johnson, and T.~G. Barraclough, 2008.
\newblock Biodiversity inhibits species' evolutionary responses to changing
  environments.
\newblock Ecology Letters 11:380--388.

\bibitem[{McPeek(2008)}]{mcpeek2008}
McPeek, M.~A., 2008.
\newblock The ecological dynamics of clade diversification and community
  assembly.
\newblock The American Naturalist 172:E270--E284.

\bibitem[{Metz et~al.(1992)Metz, Nisbet, and Geritz}]{metz_etal1992}
Metz, J. A.~J., R.~M. Nisbet, and S.~A.~H. Geritz, 1992.
\newblock How should we define fitness for general ecological scenarios.
\newblock Trends in Ecology \& Evolution 7:198--202.

\bibitem[{Nuismer and Harmon(2015)}]{nuismer_harmon2015}
Nuismer, S.~L. and L.~J. Harmon, 2015.
\newblock Predicting rates of interspecific interaction from phylogenetic
  trees.
\newblock Ecology Letters 18:17--27.

\bibitem[{Pennell and Harmon(2013)}]{pennell_harmon2013}
Pennell, M.~W. and L.~J. Harmon, 2013.
\newblock An integrative view of phylogenetic comparative methods: connections
  to population genetics, community ecology, and paleobiology.
\newblock Annals of the New York Academy of Sciences 1289:90--105.

\bibitem[{Pennell et~al.(2014)Pennell, Harmon, and Uyeda}]{pennell_etal2014}
Pennell, M.~W., L.~J. Harmon, and J.~C. Uyeda, 2014.
\newblock Is there room for punctuated equilibrium in macroevolution?
\newblock Trends in Ecology and Evolution 29:23--32.

\bibitem[{Rabosky and Hurlbert(2015)}]{rabosky_hurlbert2015}
Rabosky, D.~L. and A.~H. Hurlbert, 2015.
\newblock Species richness at continental scales is dominated by ecological
  limits.
\newblock The American Naturalist 185:572--583.
\newblock \urlprefix\url{http://dx.doi.org/10.1086/680850}.

\bibitem[{Rabosky and Lovette(2008)}]{rabosky_lovette2008}
Rabosky, D.~L. and I.~J. Lovette, 2008.
\newblock Density-dependent diversification in north american wood warblers.
\newblock Proceedings of the Royal Society, London B 275:2363--2371.

\bibitem[{Rosenzweig(1995)}]{rosenzweig1995}
Rosenzweig, M.~L., 1995.
\newblock Species Diversity In Space and Time.
\newblock Cambridge University Press, Cambridge.

\bibitem[{Rosindell et~al.(2015)Rosindell, Harmon, and
  Etienne}]{rosindell_etal2015}
Rosindell, J., L.~J. Harmon, and R.~S. Etienne, 2015.
\newblock Unifying ecology and macroevolution with individual-based theory.
\newblock Ecology Letters 18:472--482.
\newblock \urlprefix\url{http://dx.doi.org/10.1111/ele.12430}.

\bibitem[{Schluter(2000)}]{schluter2000}
Schluter, D., 2000.
\newblock The ecology of adaptive radiation.
\newblock Oxford University Press, Oxford, UK.

\bibitem[{Seehausen(2015)}]{seehausen2015}
Seehausen, O., 2015.
\newblock Process and pattern in cichlid radiations - inferences for
  understanding unusually high rates of evolutionary diversification.
\newblock New Phytologist 207:304--312.
\newblock \urlprefix\url{http://dx.doi.org/10.1111/nph.13450}.

\bibitem[{Shtilerman et~al.(2015)Shtilerman, Kessler, and
  Shnerb}]{shtilerman_etal2015}
Shtilerman, E., D.~A. Kessler, and N.~M. Shnerb, 2015.
\newblock Emergence of structured communities through evolutionary dynamics.
\newblock Journal of Theoretical Biology 383:138--144.

\bibitem[{Simpson(1944)}]{simpson1944}
Simpson, G.~G., 1944.
\newblock Tempo and Mode in Evolution.
\newblock Columbia University Press, New York.

\bibitem[{Slater(2015)}]{slater2015}
Slater, G.~J., 2015.
\newblock Not-so-early bursts and the dynamic nature of morphological
  diversification.
\newblock Proceedings of the National Academy of Sciences, USA 112:3595--3596.

\bibitem[{Slater and Pennell(2013)}]{slater_pennell2013}
Slater, G.~J. and M.~W. Pennell, 2013.
\newblock Robust regression and posterior predictive simulation increase power
  to detect early bursts of trait evolution.
\newblock Systematic Biology 63:293--308.

\bibitem[{Svardal et~al.(2014)Svardal, Rueffler, and
  Doebeli}]{svardal_etal2014}
Svardal, H., C.~Rueffler, and M.~Doebeli, 2014.
\newblock {Organismal complexity and the potential for evolutionary
  diversification}.
\newblock Evolution 68:3248--3259.
\newblock \urlprefix\url{http://dx.doi.org/10.1111/evo.12492}.

\bibitem[{Uyeda et~al.(2011)Uyeda, Hansen, Arnold, and
  Pienaar}]{uyeda_etal2011}
Uyeda, J.~C., T.~F. Hansen, S.~J. Arnold, and J.~Pienaar, 2011.
\newblock The million-year wait for macroevolutionary bursts.
\newblock Proceedings of the National Academy of Sciences, USA
  108:15908--15913.

\bibitem[{Weir and Mursleen(2013)}]{weir_mursleen2013}
Weir, J.~T. and S.~Mursleen, 2013.
\newblock Diversity-dependent cladogenesis and trait evolution in the adaptive
  radiation of the auks (aves: Alcidae).
\newblock Evolution 67:403--416.

\bibitem[{Yoon et~al.(2012)Yoon, Han, and Jeong}]{yoon_etal2012}
Yoon, S.~H., M.~J. Han, and H.~{\it et al}. Jeong, 2012.
\newblock Comparative multi-omics systems analysis of {\it escherichia coli}
  strains b and k-12.
\newblock Genome Biology 13:R37.

\end{thebibliography}
\bibliographystyle{evolution}

\end{document}